\documentclass[aps,prc,a4paper,nofootinbib,
preprintnumbers,superscriptaddress,twocolumn]{revtex4}

\usepackage{amsmath}
\usepackage{amssymb}
\usepackage{bm}
\usepackage{graphicx}
\usepackage{color}

\begin{document}

\title{Evolution of pressures and correlations in the Glasma produced in high energy nuclear collisions}

\author{M. Ruggieri}\email{ruggieri@lzu.edu.cn}
\affiliation{School of Nuclear Science and Technology, Lanzhou University, 222 South Tianshui Road, Lanzhou 730000, China.}

\author{J. H. Liu}
\affiliation{School of Nuclear Science and Technology, Lanzhou University, 222 South Tianshui Road, Lanzhou 730000, China.}

\author{L. Oliva}
\affiliation{Department of Physics and Astronomy, University of Catania, Via S. Sofia 64, I-95125 Catania}
\affiliation{INFN-Laboratori Nazionali del Sud, Via S. Sofia 62, I-95123 Catania, Italy.}

\author{G. X. Peng}
\affiliation{College of Physics, University of Chinese Academy of Sciences, Yuquanlu 19A, Beijing 100049, China.}
\affiliation{Theoretical Physics Center for Science Facilities, Institute of High Energy Physics, Beijing 100049, China.}

\author{V. Greco}
\affiliation{Department of Physics and Astronomy, University of Catania, Via S. Sofia 64, I-95125 Catania.}
\affiliation{INFN-Laboratori Nazionali del Sud, Via S. Sofia 62, I-95123 Catania, Italy.}
\affiliation{College of Physics, University of Chinese Academy of Sciences, Yuquanlu 19A, Beijing 100049, China.}


\begin{abstract}
We consider the SU(2) Glasma with gaussian fluctuations 
and study its evolution by means of classical Yang-Mills equations solved numerically on a lattice.  
Neglecting in this first study the longitudinal expansion we follow the evolution of the pressures of the system 
and compute the effect of the fluctuations in the early stage up to $t\approx 2$ fm/c, that is the
time range in which the Glasma is relevant for high energy collisions. 
We measure the ratio of the longitudinal over the
transverse pressure, $P_L/P_T$, and we find that unless the fluctuations carry a substantial amount of the energy
density at the initial time, they do not change significantly the evolution of $P_L/P_T$ in the early stage,
and that the system remains quite anisotropic. We also measure the 
longitudinal fields correlators both in the transverse plane and along the longitudinal direction: 
while at initial time 
fields appear to be anticorrelated in the transverse plane,
this anticorrelation disappears in the very early stage 
and the correlation length
in the transverse plane increases. On the other hand, we find a dependence of the gauge invariant correlator on the longitudinal
coordinate which we interpret as a partial loss of correlation induced by the dynamics, that we dub the gauge invariant string
breaking.
We finally study the effect of fluctuations on the longitudinal correlations: we find that string breaking
is accelerated by the fluctuations and waiting for a sufficiently long time fluctuations lead to
the complete breaking of the color strings.
\end{abstract}




\maketitle

\section{Introduction}

The study of the initial condition of the system produced by high energy collisions and of its evolution
to a quark-gluon plasma (QGP) is one of the most difficult but interesting problems related to the physics of
relativistic heavy ion collisions (RHICs), as well as to that of high energy proton-proton (pp) and 
proton-nucleus (pA) collisions. If the energy of the collision is very large then the two colliding nuclei
in the backward light cone can be described within the color-glass-condensate (CGC) effective theory 
\cite{McLerran:1993ni,McLerran:1993ka,McLerran:1994vd},
see~\cite{Gelis:2010nm,Iancu:2003xm,McLerran:2008es,Gelis:2012ri} for reviews:
because of Lorentz contraction along the flight direction the nuclei appear as thin sheets of a colored glass,
in which the fast partons dynamics is frozen by time dilatation and these degrees of freedom act as static sources
for low momentum gluons: their large occupation number allows to treat them as classical fields, in particular as 
transverse color-electric and color-magnetic fields.  
 
Immediately after the collision the situation changes quite drastically. As a matter of fact,
the solution of the classical Yang-Mills (CYM) equations in the forward light cone shows that
new fields are 
formed \cite{Kovner:1995ja,Kovner:1995ts,Gyulassy:1997vt,Lappi:2006fp,Fukushima:2006ax,
Fries:2006pv,Chen:2015wia,Fujii:2008km,Krasnitz:2000gz,Krasnitz:2003jw,Krasnitz:2001qu}, 
because of the non abelian interaction of the two CGC
sheets: these fields are called the 
Glasma \cite{Lappi:2006fp} and are characterized by
having, at initial time, nonvanishing longitudinal components 
while the transverse components are zero.
The Glasma fields connect the effective color charges 
that are formed on the transverse plane of the
two colliding nuclei after the collision, 
which are opposite on the two light cones $x^0 = \pm x^3$ as it should be
in order to generate longitudinal fields  \cite{Lappi:2006fp}.
These fields are classical since they are characterized by a large gluon occupation number,
typically $A_\mu^a \simeq 1/g$ with $g$ the QCD coupling.
The classical theory that describes the evolution of these fields is named the Classical Yang-Mills  (CYM) theory.
 
On the top of the Glasma it is possible to add 
quantum fluctuations \cite{Romatschke:2005pm,Romatschke:2006nk,Fukushima:2011nq,
Fukushima:2013dma,Iida:2014wea,Gelis:2013rba,Epelbaum:2013waa,Tanji:2011di,Ryblewski:2013eja,Ruggieri:2015yea,
Berges:2012cj,Berges:2013fga,Berges:2013lsa,Berges:2013eia} 
which are known to trigger plasma instabilities and are helpful to produce entropy during the early stage 
of high energy nuclear collisions. 
Quantum fluctuations appear when one considers the finite coupling corrections to the Glasma solution which is obtained
in the small coupling limit;  the spectrum of these fluctuations has been computed within a perturbative calculation in~\cite{Gelis:2013rba}
and it has been shown that they affect both the gauge potential and the color electric field.
Although quantum in nature, it is interesting to study what are the predictions of the CYM theory
on the evolution of the system made of a Glasma plus the fluctuations, although the application of the classical theory
is justified only if the amount of fluctuations is small in comparison with the background classical field.
In this article we will follow \cite{Romatschke:2005pm,Romatschke:2006nk,Iida:2014wea,Berges:2012cj} 
and we will consider for the sake of simplicity
the fluctuations of the color electric field only, neglecting those of the gauge potentials
that will be considered in future works.

We present here some result on the role of classical fluctuations on the isotropization of Glasma.
The quantity that we will consider for this problem is the ratio of the longitudinal over the 
transverse pressures, $P_L/P_T$, which has been the subject of studies on the onset of the 
hydrodynamical regime in the early stage of relativistic nuclear collisions 
\cite{Romatschke:2005pm,Romatschke:2006nk,Fukushima:2011nq,Gelis:2013rba,
Ryblewski:2013eja,Ruggieri:2015yea,Li:2016eqr,Kurkela:2015qoa,Bellantuono:2015hxa}.
We consider here the Glasma in a three-dimensional static box with periodic boundary conditions, neglecting for simplicity the
longitudinal expansion that will be considered in a next work: the longitudinal expansion is necessary
to describe the early stage of the system created in realistic collisions but focusing on the static box
allows for a clear intepretation of the physical results without the complication of the dilution of the fields.

We will follow closely the work of \cite{Iida:2014wea} in which gaussian fluctuations have been studied
in relation to entropy production, but almost no emphasis has been put on isotropization.
Moreover, our work has some link to \cite{Fukushima:2013dma} in which a similar problem has been considered
but within a simplified model of fluctuations. 
Instead of insisting to study the evolution of the system up to very late times as in
\cite{Fukushima:2013dma,Iida:2014wea} we will focus on the very early stage,
which is the only time range in which the description based on CYM
has some phenomenological interest for high energy nuclear collisions.  

We can anticipate one of our results, namely that
as long as fluctuations carry a not so large part of the initial energy of the Glasma, their impact on the
evolution of the longitudinal over transverse pressure is marginal in the early stage.
This result will hardly be modified by the longitudinal expansion because it is known that the latter
can only lower the amount of isotropy as well as delay the isotropization \cite{Ruggieri:2015yea}. 
We also find that increasing the amount of energy carried by the fluctuations it is possible to obtain
a fairly isotropic system within a short time range: most likely this is due to the saturation of the instabilities 
as mentioned in \cite{Fukushima:2013dma,
Romatschke:2005pm,Romatschke:2006nk,Fukushima:2011nq}. 
Due to our limitated computing power we do not perform long time running simulations here,
therefore we are not able to put a firm statement on the fact that instabilities rather than a collisional dynamics lead to
the quick isotropization in cases with substantial fluctuations: in this article we limit ourselves to present few results
about the isotropization (or missed isotropization) in the early stage and how the fluctuation size affects the evolution
of $P_L/P_T$, leaving more specific studies to a near future work. 

Besides the pressures there are other quantities that are interesting to understand
the evolution of the color fields produced in high energy nuclear collisions: correlation functions of the classical color fields
represent an example of these . The correlators have been studied in several 
articles before \cite{Fujii:2008km,Dumitru:2013wca,Dumitru:2014nka} and it has been shown that 
at $\tau=0^+$ the Glasma presents a very short correlation magnetic length, $\lambda_M Q_s \ll 1$: this means that a description 
of the initial stage in terms
of uncorrelated color strings is appropriate. 
Moreover, the investigation of the gauge invariant 
magnetic correlators shows that for $Q_s x_T \gtrsim 1$ the Glasma fields present anticorrelation, which 
roughly speaking implies
that in domains of transverse area $A_T\approx Q_s^2$ the magnetic field flips its sign \cite{Dumitru:2014nka}.
On the other hand,  for $\tau = O(1/Q_s)$ the anticorrelation disappears and the correlation length increases:
the interaction tends to align the magnetic field and to correlate the fields in domains with $A_T\approx Q_s^2$.

Another problem that we study in this article
is the impact of fluctuations on the evolution of the field correlators. 
Once again we consider only the static box case: the longitudinal expansion
can dilute the correlators but cannot affect drastically their structure in the transverse plane, that is the one
we consider here. Also in this case we can anticipate our main results, namely that 
fluctuations do not change drastically the picture described above.
Another novelty that we bring here is the study of the longitudinal correlation functions:
these are important to understand if and how the color strings experience some loss of correlation
along the longitudinal direction during the CYM evolution. We anticipate our main result on this, namely that
even in absence of fluctuations the evolving Glasma experiences a partial loss of correlation
on the long distance within a very short time. We call this as the gauge invariant partial string breaking. 
We dub this as the partial string breaking since we find some residual correlation also at large distance
for $g^2\mu t\simeq 10$, although this correlation is smaller than the one we have at a very small distance;
moreover, we specify that this is gauge invariant since it is related to the calculation of the gauge invariant
correlation function, while the naive gauge dependent correlator would be $z-$independent at any time.
We also study the effect of fluctuations on the longitudinal correlations: we find that string breaking
is accelerated by the fluctuations and waiting for a sufficiently long time fluctuations lead to
the complete breaking of the color strings.

The plan of the article is as follows: in section II, we review the initial condition that we implement in 
the calculations, as well as the spectrum of the fluctuations and the CYM equations that we solve to study the
evolution of the Glasma. In section III, we present results about pressures evolution with and without 
fluctuations. In section IV, we discuss the correlation functions of the Glasma fields in the transverse plane.
Finally, in section V we draw our conclusions.

\section{Glasma and classical Yang-Mills equations}
In this section we briefly review how the Glasma is built up within the McLerran-Venugopalan (MV) 
model \cite{McLerran:1993ni,McLerran:1993ka,McLerran:1994vd,Kovchegov:1996ty} and gaussian fluctuations
are added on the top of it,
then how this initial condition is evolved by means of the CYM equations.

\subsection{The Glasma\label{sec:glasma}}
In the MV model, the color charge densities $\rho_a$ that act like the static sources of the CGC fields 
in the two colliding nuclei are assumed to be random variables that are 
normally distributed with zero mean 
and variance specified by the equation
\begin{equation}
\langle \rho^a(\bm x_T)\rho^b(\bm y_T)\rangle = (g^2\mu)^2 \delta^{ab}\delta^{(2)}(\bm x_T-\bm y_T);
\label{eq:dfg}
\end{equation}
$a$ and $b$ denote the adjoint color index;
in this work we limit ourselves for simplicity to the case of the $SU(2)$ color group therefore
$a,b=1,2,3$.
In Eq.~(\ref{eq:dfg}) $g^2\mu$ is the only energy scale of the model that is related to the saturation momentum $Q_s$:
lattice calculations of the Wilson line correlations for deep inelastic scattering show that 
$Q_s / g^2\mu$ in the MV model lies approximately in the range (0.6,1.2) \cite{Lappi:2007ku},
while for realistic high energy nuclear collisions the relation between the two quantities is less clear
because one has to sum up over gluons produced with different transverse momenta, hence related
to partons with different values of $x$ in the nuclear wave function; in this case 
$Q_s/g^2\mu\approx 0.6$ for nuclear collisions at the RHIC energy \cite{Lappi:2007ku}.
Because of the $\delta^{(2)}(\bm x_T-\bm y_T)$ the fluctuations of the color charge 
density in the two CGC sheets are uncorrelated in the transverse plane. 

On the lattice Eq.~(\ref{eq:dfg}) is implemented by distributing the fluctuating color charges with
variance given by  
\begin{equation}
\langle \rho^a(\bm x_T)\rho^b(\bm y_T)\rangle = 
(g^2\mu)^2 \delta^{ab}
\frac{1}{a^2}\delta_{\bm x_T,\bm y_T},
\label{eq:poi1}
\end{equation}
where $a\equiv L_x/N_x=L_y/N_y$ is the lattice spacing, with
$L_x$, $L_y$ corresponding to the physical length of the lattice in the $x$ and $y$ directions respectively,
while $N_x$, $N_y$ denote the number of cells in the $x$ and $y$ directions;
in this work we assume that $L_x = L_y$  and $N_x = N_y$. In the lattice implementation we remove the zero mode
from the color charge density: physically this amounts to require that the net color charge carried by the
distribution is vanishing; this is achieved by Fourier transforming
$\rho_a(\bm x_T)$ for each $a$, then constructing a new density $\tilde\rho_a(\bm x_T)$
summing over all the Fourier modes but the zero mode.
For the sake of notation we denote in the following by $\rho_a(\bm x_T)$ the density obtained in this way,
keeping in mind that it corresponds to a charge distribution that has been color neutralized.

The static color sources $\{\rho\}$ generate CGC fields that can be computed as follows.
Firstly we  solve the Poisson equations for the gauge potentials
generated by the color charge distributions of the nuclei $A$ and $B$, namely
\begin{equation}
-\partial_\perp^2 \Lambda^{(A)}(\bm x_T) = \rho^{(A)}(\bm x_T)
\end{equation}
(a similar equation holds for the distribution belonging to $B$). The Wilson lines are then computed as
\begin{equation}
V^\dagger(\bm x_T) = e^{i \Lambda^{(A)}(\bm x_T)},~~~
W^\dagger(\bm x_T) = e^{i \Lambda^{(B)}(\bm x_T)},
\end{equation}
and the pure gauge fields of the two colliding nuclei are given by
\begin{equation}
\alpha_i^{(A)} = i V \partial_i V^\dagger,~~~\alpha_i^{(B)} = i W \partial_i W^\dagger.
\label{eq:pgp}
\end{equation}
In terms of these fields the solution of the CYM in the forward light cone
at initial time, namely the Glasma gauge potential, 
can be written as \cite{Kovner:1995ja,Kovner:1995ts} 
\begin{eqnarray}
&&A_i = \alpha_i^{(A)} + \alpha_i^{(B)}~,~~i=x,y, \\
&&A_z = 0,
\end{eqnarray}
and the initial longitudinal Glasma fields are
\begin{eqnarray}
&& E^z = i\sum_{i=x,y}\left[\alpha_i^{(B)},\alpha_i^{(A)}\right], \label{eq:f1}\\
&& B^z = i\left(
\left[\alpha_x^{(B)},\alpha_y^{(A)}\right]  + \left[\alpha_x^{(A)},\alpha_y^{(B)}\right]  
\right),\label{eq:f2}
\end{eqnarray}
while the transverse fields are vanishing\footnote{
It is worth mentioning that in the more correct implementation of the MV model the gauge potentials are 
computed as path ordered
exponentials of multiple layers of color charges, describing the propagation
of a colored probe through a thick nucleus 
\cite{Lappi:2007ku}, while in the present work we limit ourselves to consider a single layer of charges.
We will implement the sources with a finite thickness in our future works.}.

On the top of the Glasma specified by Eqs.~(\ref{eq:f1}) and~(\ref{eq:f2}) it is possible to add gaussian 
fluctuations \cite{Romatschke:2005pm,Romatschke:2006nk,Fukushima:2011nq,Iida:2014wea}
which are known to trigger instabilities and might be helpful to achieve isotropization
as well as to produce entropy during the early stage of high energy nuclear collisions;
in this article we study the Glasma in a static box therefore we use
the definitions of \cite{Iida:2014wea}. Below we report the definitions of the longitudinal and transverse
fluctuations in the continuum limit: the ones on the lattice can be obtained by extracting the trivial scaling
of the physical quantities with the proper power of the lattice spacing $a$ and the derivatives with
central difference operators.  
Firstly, we define the auxiliary random fields $\{\xi_i\}$, $i=x,y$, that are assumed to be white noise with variance
\begin{equation}
\langle\xi_i^a(\bm x_T)\xi_j^b(\bm y_T)\rangle = \delta_{ab}\delta_{ij}\delta^{(2)}(\bm x_T - \bm y_T);
\label{eq:xi1}
\end{equation}
in addition to these we consider a random fluctuation in the longitudinal direction, again normally distributed with
variance
\begin{equation}
g^2\mu \langle F(z)F(z^\prime)\rangle = \Delta^{2}\delta(z - z^\prime).
\label{eq:xi2}
\end{equation}
We notice that the $g^2\mu$ on the left hand side of the above equation has been introduced
in \cite{Iida:2014wea} to balance the inverse length dimension carried by the $\delta(z - z^\prime)$ on the
right hand side: this is necessary in the case of the static three dimensional box but it does not
appear the case of the expanding geometry \cite{Romatschke:2006nk,Fukushima:2011nq}
where $\delta(\eta - \eta^\prime)$, with $\eta$ corresponding to the space-time rapidity, 
appears instead of $\delta(z - z^\prime)$. 
The fluctuations of the electric field are then defined in the continuum limit as
\begin{eqnarray}
\delta E_i^a (\bm x_T,z) &=& \partial_z F(z) \xi_i^a(\bm x_T), \label{eq:f11}\\
\delta E_z^a (\bm x_T,z) &=& -F(z) D_i \xi_i^a(\bm x_T).\label{eq:f12}
\end{eqnarray}
These fluctuations allow to store part of the initial energy in the transverse
color electric field, as well as to break the longitudinal invariance (the boost invariance
in the case of the expanding geometry) that characterizes the Glasma fields in Eq.~(\ref{eq:f1}).

\subsection{Classical Yang-Mills equations}
In this subsection we describe how we study
the evolution of the initial conditions specified above. 
The dynamical evolution that we study here is given by the 
classical Yang-Mills (CYM) equations. 
In this study we closely follow the work of \cite{Iida:2014wea} therefore we refer
to that reference for more details; 
we write below the relevant equations in lattice units therefore the 
physical quantities do not carry any mass dimension. 

The CYM hamiltonian density is given by
\begin{equation}
H = \frac{1}{2}\sum_{x,a,i}E_i^a(x)^2 + \frac{1}{4}\sum_{x,a,i,j}F_{ij}^a(x)^2,
\label{eq:H} 
\end{equation}
where the magnetic part of the field strength tensor is  
\begin{equation}
F_{ij}^a(x) = \partial_i A_j^a(x) - \partial_j A_i^a(x)  - \sum_{b,c}f^{abc} A_i^b(x) A_j^c(x);
\label{eq:Fij}
\end{equation}
here $\partial_i A_j^a(x) = (A_j(x+i)-A_j(x-i))/2$ corresponds to the central difference
operator in the $i^{\mathrm{th}}$-direction, $f^{abc} = \varepsilon^{abc}$ with $\varepsilon^{123} = +1$.
The equations of motion for the fields and conjugate momenta, namely the CYM equations, are 
\begin{eqnarray}
\frac{dA_i^a(x)}{dt} &=& E_i^a(x),\\
\frac{dE_i^a(x)}{dt} &=& \sum_j \partial_j F_{ji}^a(x) - 
\sum_{b,c,j} f^{abc} A_j^b(x)  F_{ji}^c(x).
\end{eqnarray}
We solve the above equations on a static box in three spatial dimension, using the common 
fourth order Runge-Kutta method as in \cite{Iida:2014wea}.
Although the method used in this work sounds a bit primitive in comparison with 
recent works that formulate the problem in terms of gauge links,
see for example \cite{Fukushima:2011nq,Gelis:2013rba}, 
the approach used here gives results that are in good agreement with these methods and
we leave the more rigorous implementation of the numerical problem based on gauge links
to a future work.

\section{Results: fields and pressures\label{sec:ruv}}
In this section we show our results obtained for the evolution of the Glasma in the case of a static box
with a square cross section of side $L$.
Our main goal is to study the color fields produced in high energy nuclear collisions,
although within a simplified static geometry: 
we set the physical length of the box in the transverse plane  to $L = 2$ fm; moreover, 
we consider $g^2\mu=1$ GeV.

We notice that although it would be possible to extend the numerical calculations in the static box up to
$g^2\mu t \gg 1$, in the real collisions the longitudinal expansion would dilute the fields thus invalidating the
classical approximation used here: previous studies with the expanding geometry show that dilution of the fields
becomes important for  $g^2\mu t =O(1)$, therefore for times
$g^2\mu t\gtrsim O(1)$ particle quanta dynamically evolving in interaction with  the classical fields should be 
introduced, see for example \cite{Gelis:2013rba,Ryblewski:2013eja,Ruggieri:2015yea}. 
For this reason in this study we will focus to times ranging up to 
few units of $1/g^2\mu$, that in physical units correspond to $t\approx 2$ fm/c. 

\subsection{Fields}

\begin{figure}[t!]
\begin{center}
\includegraphics[scale=0.3]{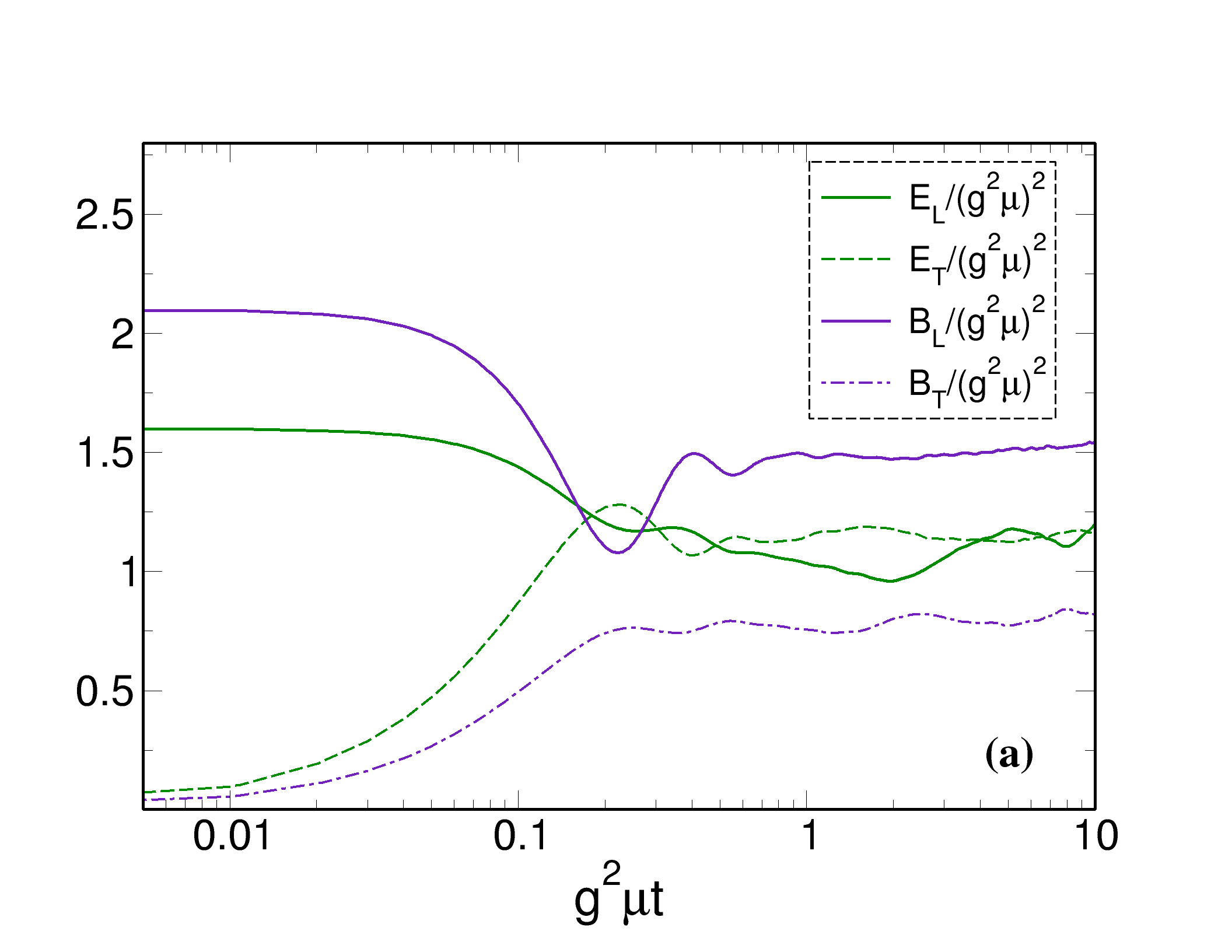}\\
\includegraphics[scale=0.3]{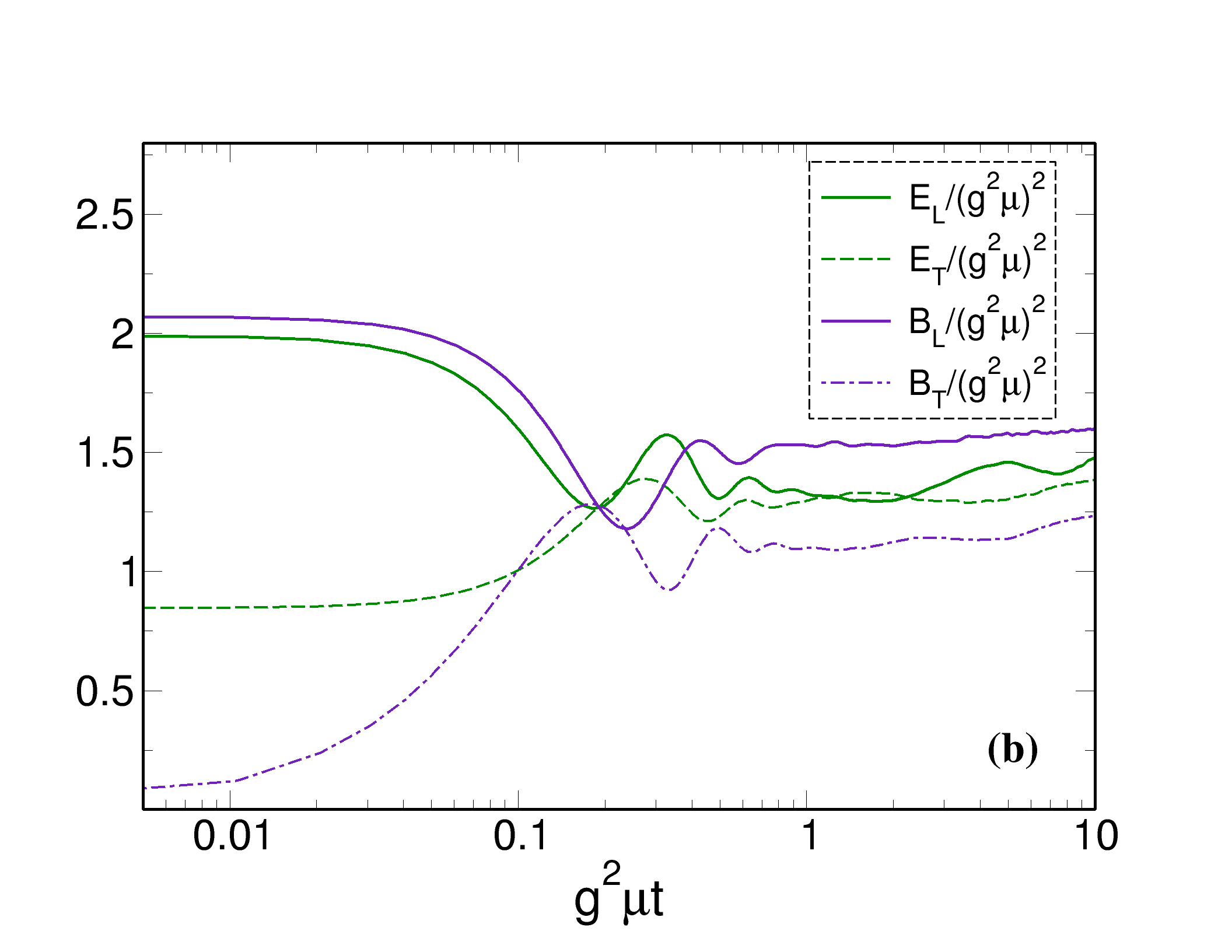}
\end{center}
\caption{\label{Fig:1}Glasma fields, averaged over the whole box, as a function of time measured in units of $g^2\mu$.
All the fields are measured in units of $(g^2\mu)^2$. Indigo lines correspond to the magnetic color fields and green lines
to the electric color fields. Moreover, solid lines denote longitudinal fields while dashed lines correspond to the
averaged transverse fields. 
Upper panel stands for the case without fluctuations, while the lower panel collects the results obtained 
adding gaussian fluctuations with the seed $\Delta^2=2.2\times 10^{-4} g^2\mu a$.
The results correspond to a grid with $N=71^3$ lattice sites with lattice spacing $a$ equal to
$g^2\mu a = 0.14$. }
\end{figure}

In Fig.~\ref{Fig:1} we plot the Glasma fields, averaged over the whole box, as a function of time measured in units 
of $g^2\mu$.
All the fields are measured in units of $(g^2\mu)^2$. Indigo lines correspond to the magnetic color fields and green lines
to the electric color fields. Moreover, solid lines denote longitudinal fields while dashed lines correspond to the
averaged transverse fields. 
The upper panel stands for the case without fluctuations, while the lower panel collects the results obtained 
adding gaussian fluctuations with the seed $\Delta^2=2.2\times 10^{-4} g^2\mu a$.
We have averaged each component summing over color indices building up gauge invariant
quantities: for example we have defined
\begin{equation}
B_L^2 = \frac{1}{N}\sum_{i,j=-N_x}^{N_x}\sum_{k=-N_z}^{N_z}
\sum_{a=1}^3 \left[B_z^a(x_i,y_j,z_k)\right]^2,
\end{equation}
and
\begin{eqnarray}
B_T^2 &=&\frac{1}{2}\left( \frac{1}{N}\sum_{i,j=-N_x}^{N_x}\sum_{k=-N_z}^{N_z}
\sum_{a=1}^3 
\left[B_x^a(x_i,y_j,z_k)\right]^2
\right.\nonumber\\
&& +\left.\frac{1}{N}\sum_{i,j=-N_x}^{N_x}\sum_{k=-N_z}^{N_z}
\sum_{a=1}^3  \left[B_y^a(x_i,y_j,z_k)\right]^2\right);
\end{eqnarray}
similar definitions hold for the color electric fields. In the above equations $N$ corresponds to the number
of lattice sites $N=(2N_x+1)\times (2N_y+1) \times (2N_z+1)$ with $N_x = N_y = N_z = 35$.

The results collected in the upper panel of Fig.~\ref{Fig:1} show that at the initial time
the system is made of purely longitudinal fields; however, in a time range $g^2\mu t \approx 0.5$ the bulk structure
of the fields is formed, with both longitudinal and transverse fields having reached their asymptotic value: 
in fact going on with time the evolution does not change drastically the value of the fields. 

In the lower panel of Fig.~\ref{Fig:1}  we show the evolution of the classical fields with gaussian fluctuations in the
initial condition. We notice the main effect of the fluctuations on the averaged fields is to have nonvanishing
transverse electric fields at the initial time. Besides this the evolution of the fields follows that measured in case
of the pure Glasma, with the bulk structure of the fields that is formed within $g^2\mu t \approx 0.5$.

\subsection{Pressures evolution}

For a system made of only classical color fields the longitudinal and transverse pressures are given by
\begin{eqnarray}
P_T &=&\frac{E_z^a E_z^a + B_z^a B_z^a}{2},\label{eq:uyt1}\\
P_L &=&-P_T+ 
  \frac{\bm{E}_T^a \bm{E}_T^a +\bm{B}_T^a \bm{B}_T^a}{2},
  \label{eq:uyt2}
\end{eqnarray}
with $\bm{E}_T^a \bm{E}_T^a = E_x^a E_x^a + E_y^a E_y^a$,
$\bm{B}_T^a \bm{B}_T^a = B_x^a B_x^a + B_y^a B_y^a$; isotropy requires $P_L = P_T$.

At initial time $P_L/P_T=-1$ in the pure Glasma, see also below, and a hydro description 
of the fireball created in relativistic collisions would prefer that a nearly isotropic system is formed within a short time 
range $\approx 1$ fm/c: 
\footnote{A perfect isotropy is however not necessary, in fact it has been realized that
anisotropic hydro can succesfully describe the evolution of the quark-gluon plasma 
even if at the initialization time the isotropy is not perfect, see 
\cite{Martinez:2012tu,Martinez:2010sd,Strickland:2016ezq,Alqahtani:2017tnq,
Florkowski:2010cf,Florkowski:2013lza,Florkowski:2013lya,Ryblewski:2012rr} and
references therein.}
it is therefore interesting to address the question whether the classical evolution of the Glasma can
produce a fairly isotropic system within this short amount of time.  

\begin{figure}[t!]
\begin{center}
\includegraphics[scale=0.3]{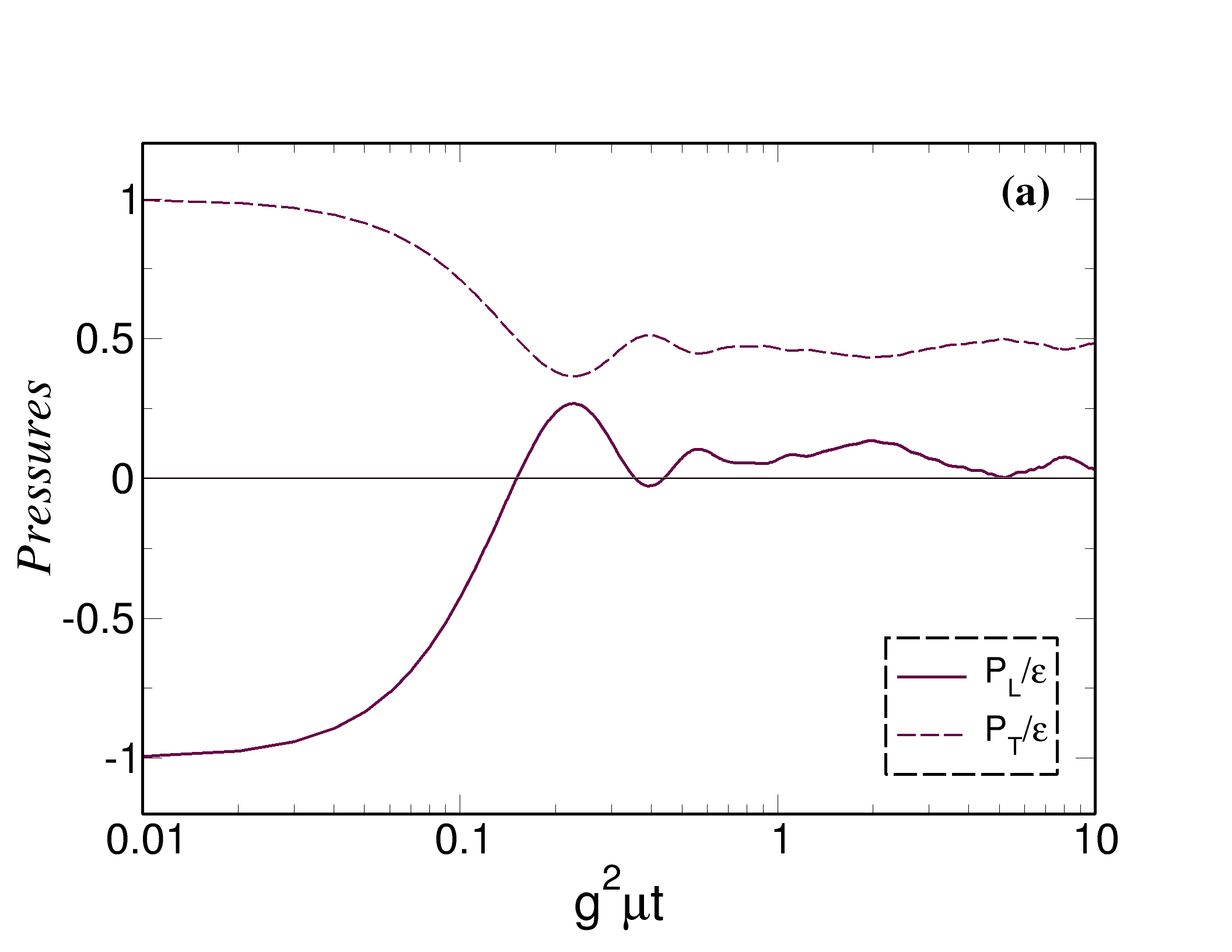}\\
\includegraphics[scale=0.3]{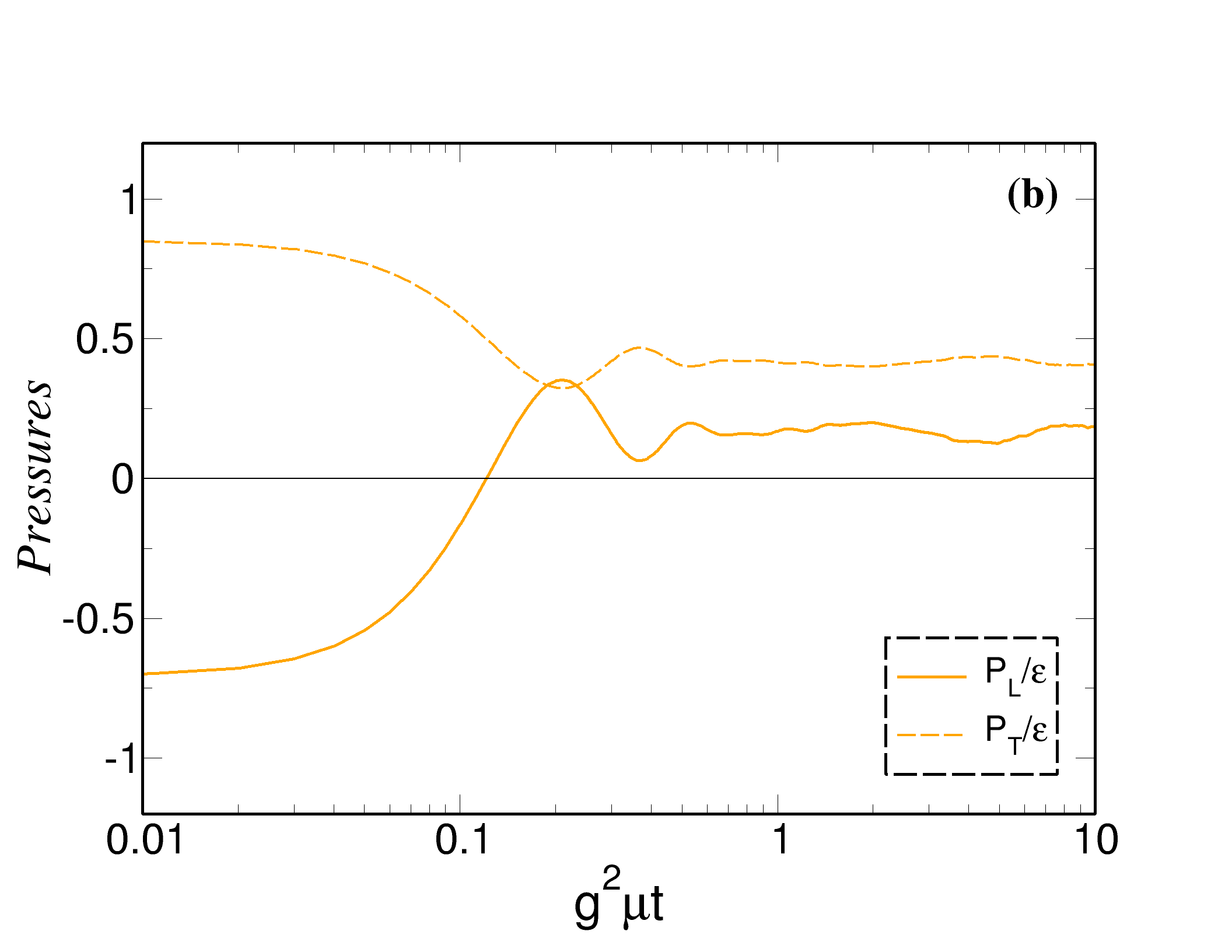}
\end{center}
\caption{\label{Fig:3before}Pressures over energy density as a function of $g^2\mu t$.
Maroon lines in panel {\bf{(a)}} correspond to $P_L/\varepsilon$ (solid) and $P_T/\varepsilon$ (dashed) for the case without fluctuations;
orange lines  in panel {\bf{(b)}} correspond to $P_L/\varepsilon$ (dot-dashed) and $P_T/\varepsilon$ (dotted)
for the case with gaussian fluctuations with $\Delta^2 = 2.2\times 10^{-4}g^2\mu a$.
In both panels the lattice specifications as well as the value of $g^2\mu$ are the ones used for the 
data shown in Fig.~\ref{Fig:1}.}
\end{figure}

\begin{figure}[t!]
\begin{center}
\includegraphics[scale=0.3]{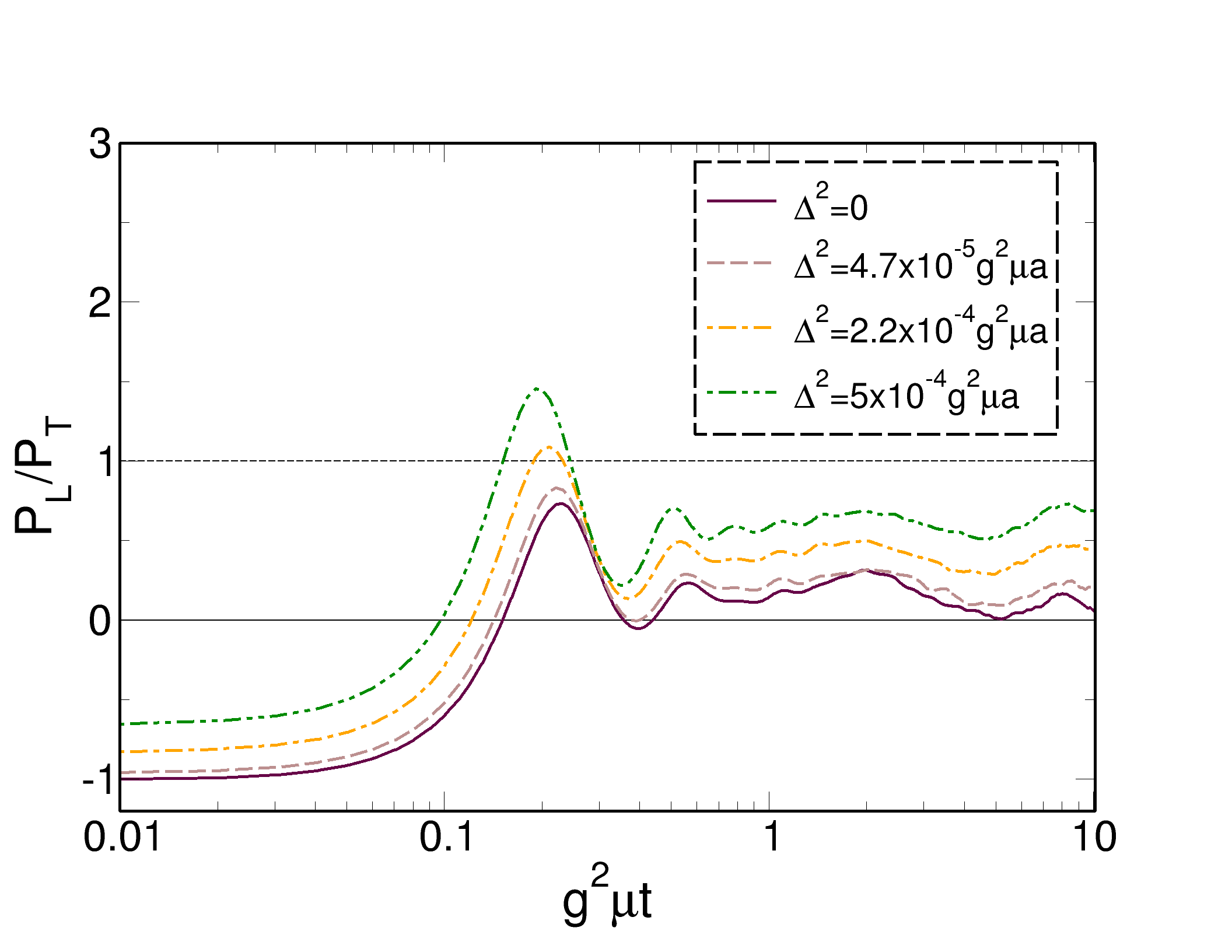}
\end{center}
\caption{\label{Fig:3}{\em Upper panel.} Pressures as a function of $g^2\mu t$.
Maroon lines correspond to $P_L/\varepsilon$ (solid) and $P_T/\varepsilon$ (dashed) for the case without fluctuations;
orange lines correspond to $P_L/\varepsilon$ (dot-dashed) and $P_T/\varepsilon$ (dotted)
for the case with gaussian fluctuations with $\Delta^2 = 2.2\times 10^{-4}g^2\mu a$.
{\em Lower panel.} Ratio $P_L/P_T$ for several values of $\Delta^2$.
In both panels the lattice specifications as well as the value of $g^2\mu$ are the ones used for the 
data shown in Fig.~\ref{Fig:1}.}
\end{figure}

In the upper panel of Fig.~\ref{Fig:3} we plot the
pressures as a function of $g^2\mu t$ for the cases of the pure longitudinal initial fields,
while in the lower panel we show the results for the case of gaussian fluctuations with $\Delta^2/g^2\mu a = 2.2\times 10^{-4}$
analogously to the data shown in  Fig.~\ref{Fig:1}. 
In the figure, solid lines correspond to $P_L/\varepsilon$ and dashed lines to $P_T/\varepsilon$.

We firstly comment on the case $\Delta=0$.
At initial time the Glasma consists of purely longitudinal fields therefore
$P_L/P_T = -1$, see Eq.~(\ref{eq:uyt2}), and the system presents
a strong anisotropy. However, the Yang-Mills evolution tends to remove part of this initial anisotropy by making
the longitudinal pressure positive for $g^2\mu t \approx 0.15$. On the other hand, the interactions are not enough to
isotropize the system, indeed at later times a strong anisotropy remains with $P_L\ll P_T$
in agreement with the static box case studied in \cite{Fukushima:2013dma}
and similarly to what happens in case of a longitudinally expanding 
geometry \cite{Fukushima:2011nq,Gelis:2013rba,Li:2016eqr}.

Turning on the gaussian fluctuations results in the increase of the 
initial value of $P_L$, thanks to the presence of transverse electric fields in the initial state.
For the case $\Delta^2=2.2\times 10^{-4} g^2\mu a$ shown in Fig.~\ref{Fig:3before}
we find that the disturbances approximately
add a $30\%$ to the total energy. 
We notice that fluctuations
are not enough to gain a considerable amount of isotropization within a short time range.

In Fig.~\ref{Fig:3} in which we plot $P_L/P_T$ for
several values of $\Delta$: in particular, the maroon solid line corresponds to $\Delta=0$;
the brown dashed line to $\Delta^2/g^2\mu a = 4.7\times 10^{-5}$ for which the energy carried by the fluctuations
is $\approx 10\%$ of the total energy; the dot-dashed orange line corresponds to $\Delta^2/g^2\mu a = 2.2\times 10^{-4}$
for which the energy carried by the fluctuations
is $\approx 30\%$ of the total energy; finally, the green dot-dot-dashed line denotes the case 
$\Delta^2/g^2\mu a = 5\times 10^{-4}$
for which the energy carried by the fluctuations
is $\approx 50\%$ of the total energy.
We notice that 
one of the effects of the  fluctuations is to increase the value of $P_L/P_T$
that is reached for $g^2\mu t \gtrsim 1$: in fact, for $\Delta=0$ the ratio of pressures
remains positive but below $\approx 0.3$ for almost all the range of time studied here, namely
$0.5 \lesssim g^2\mu t \lesssim 10$ that corresponds to the physical time range
$0.1~\mathrm{fm/c} \lesssim g^2\mu t \lesssim 2~\mathrm{fm/c}$.
Small fluctuations do not change significantly this result, see the brown dashed line in Fig.~\ref{Fig:3}.
For a more substantial amount of fluctuations the ratio of pressures in the aforementioned time range
is larger: for example, $0.3 \lesssim P_L/P_T \lesssim 0.5$ for $\Delta^2/g^2\mu a = 2.2\times 10^{-4}$
and $0.5 \lesssim P_L/P_T \lesssim 0.73$ for $\Delta^2/g^2\mu a = 5\times 10^{-4}$.
We thus obtain that in order to have a fairly isotropic system within a short time range
it is important to have a substantial amount of fluctuations in the initial stage.

Before going ahead we would like to mention that the results collected in Fig.~\ref{Fig:3}
lead to a picture that is consistent with the one discussed in \cite{Gelis:2013rba}. As a matter of fact,
quantum fluctuations have been considered in \cite{Gelis:2013rba} for two values of the QCD coupling, $g$, and it has been found that
for the relatively small value $g=0.1$ the fluctuations are not enough to gain a substantial amount of isotropization in the early stage,
while for the larger value $g=0.5$ the instabilities affect in a more substantial way the early evolution of $P_L/P_T$
and the final amount of isotropy grows up thanks to these instabilities. The results presented here confirm the scenario already anticipated
in \cite{Gelis:2013rba} and suggest that the early isotropization in the Glasma can be obtained via fluctuations-induced induced instabilities 
if fluctuations are large enough.
Our numerical estimates suggest that if fluctuations carry about the $30\%$ of the initial total energy then at later times
$P_L/P_T$ stays between $0.3$ and $0.5$, while pushing the initial energy fraction to $50\%$ brings the amount of isotropy
between $0.5$ and $0.73$. Calculations with different values of $g^2\mu$ and $a$ will be performed in the near future in order to establish
the dependence of the results on these quantities.
It will be interesting in the near future to use long time simulations of the classical Yang-Mills dynamics 
to explore the dependence of the isotropization times on the fluctuation seed: in the weak coupling scenario it is known that
the time scale for the excitation of the secondary instabilities, which are responsible for the rise of $P_L/P_T$ at very late
time, depends logarithmically on the seed \cite{Berges:2012cj}: it will be interesting to check whether this dependence remains also for
larger values of the seed.

\section{Correlation functions of the longitudinal fields}
In this section we compute the correlation functions of the longitudinal fields
in the evolving Glasma: these quantities are
important to understand the structure of the Glasma color strings and how this changes because of the CYM 
evolution. 
In particular, at the initial time the Glasma consists of longitudinal color strings that are (almost)
uncorrelated in the transverse plane; on the other hand, the CYM evolution is capable to increase the correlation
length in the transverse plane in a time range $\tau\approx 1/Q_s$.  
The correlation functions of the magnetic field have been already studied in \cite{Dumitru:2013wca,Dumitru:2014nka}
for the case of the pure Glasma. 
We present here the results for the magnetic as well as the electric correlations 
in the transverse plane for the cases without and with fluctuations, although we have found that the fluctuations
do not have a substantial impact on the evolution of these correlators. 
Then we present our results for the case of the correlations of the magnetic field along the longitudinal direction,
which have not been computed in \cite{Dumitru:2013wca,Dumitru:2014nka} and offer a way to discuss the string breaking
in the Glasma: again, we present both the results with and without fluctuations.
It is interesting to mention since now that this loss of correlation of the magnetic field at later times 
comes purely from the gluon dynamics, since in the model that we implement in the initial condition
the fluctuations are added only on the top of the electric field, giving an infinite magnetic correlation length
at the initial time both with and without fluctuations. 

\subsection{Correlation functions in the transverse plane}
In the continuum limit the correlator of the longitudinal magnetic field reads
\begin{eqnarray}
C_{M}(x_\perp)&=&\mathrm{Tr}\left\langle B_z(0) 
U_{0\rightarrow x}B_z(x_\perp)U^{-1}_{0\rightarrow x} \right\rangle,
\end{eqnarray}
where $x_\perp$ corresponds to distance measured in the transverse plane and 
we have defined $B_z = F_{12} = F_{12}^a T_a$. In the above equation
$U_{0\rightarrow x} =P e^{-i \int A_i dx}$ stands for the parallel transporter along a straight line that connects
the point $x$ with the point $x=0$, where $P$ denotes the path ordering.
In the continuum limit the path ordering would require to divide the path between two points
into infinitesimally small paths, then compute the matrix exponential of the integral along each of these 
small paths and multiply these in order to build up the full path.
In the numerical calculations the smallest path is of course that connecting two lattice neighbours points 
so the parallel transporter is implemented as follows:
suppose we aim to compute $U_{X_1\rightarrow X_{M}}$ along the $x$ direction between 
two lattice points. 
We then have to compute the $SU(2)$ matrix
\begin{equation}
U_{X_1\rightarrow X_{M}} = \prod_{j=1}^{M-1} e^{-i A_x(X_j)},
\label{eq:PathOrdering}
\end{equation}
where we remind $A_x$ is understood in lattice units and the argument of the exponential is a discretized 
version of the integral $\int_{X_j}^{X_{j+1}} A_x dx$. Then we
use the generalized Euler identity for an $SU(2)$ matrix to compute each of the exponentials
in the path ordered expression~(\ref{eq:PathOrdering}).
For the electric field we compute a similar correlation function
replacing $B_z$ with $E_z$.
On the lattice we define the average of the quantity $O$ as
\begin{equation}
\langle O\rangle = \frac{1}{N}\sum_{x_j} O(x_j),
\end{equation}
where $x_j$ denotes the $j^\mathrm{th}$ lattice site
and $N$ is the number of lattice sites involved in the summation.

\begin{figure}[t!]
\begin{center}
\includegraphics[scale=0.3]{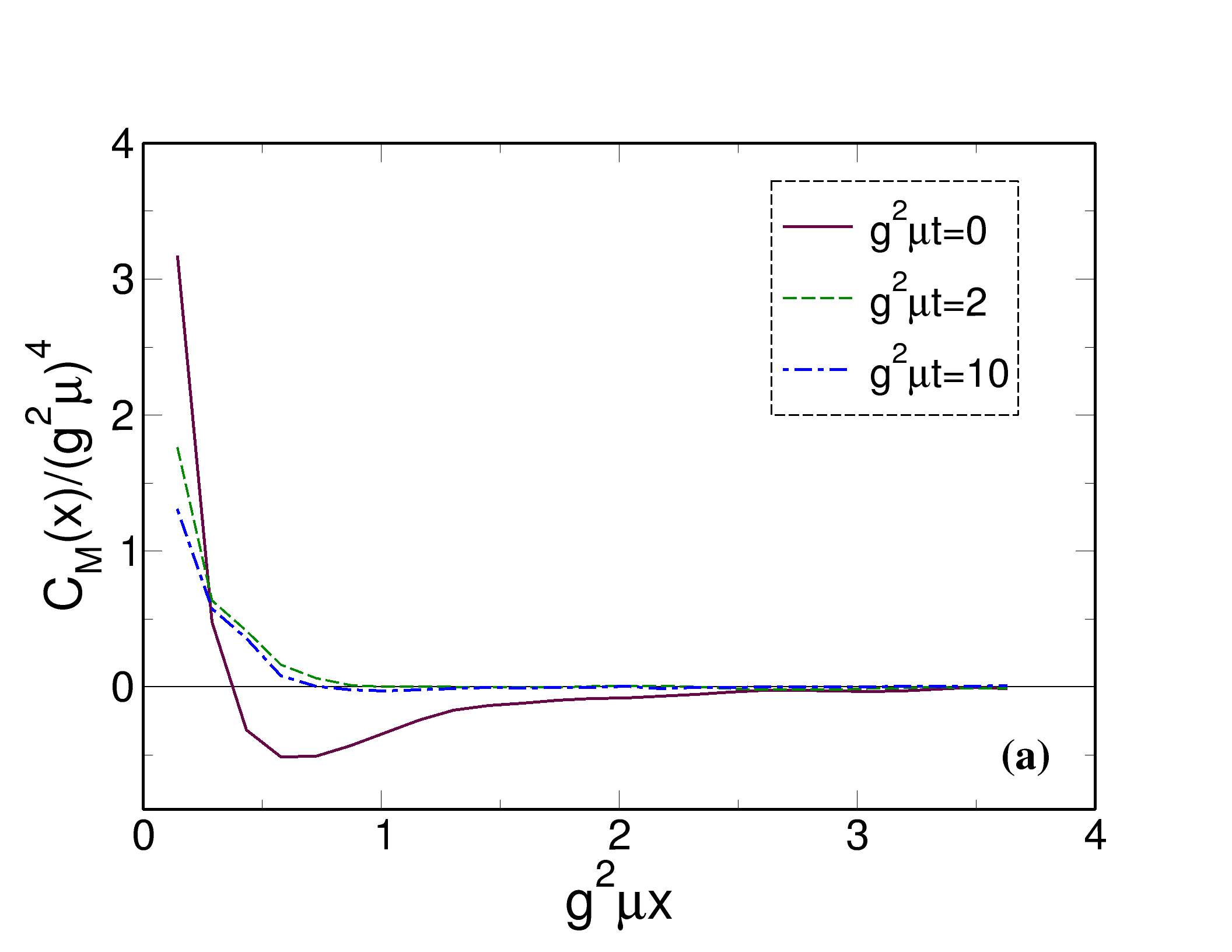}\\
\includegraphics[scale=0.3]{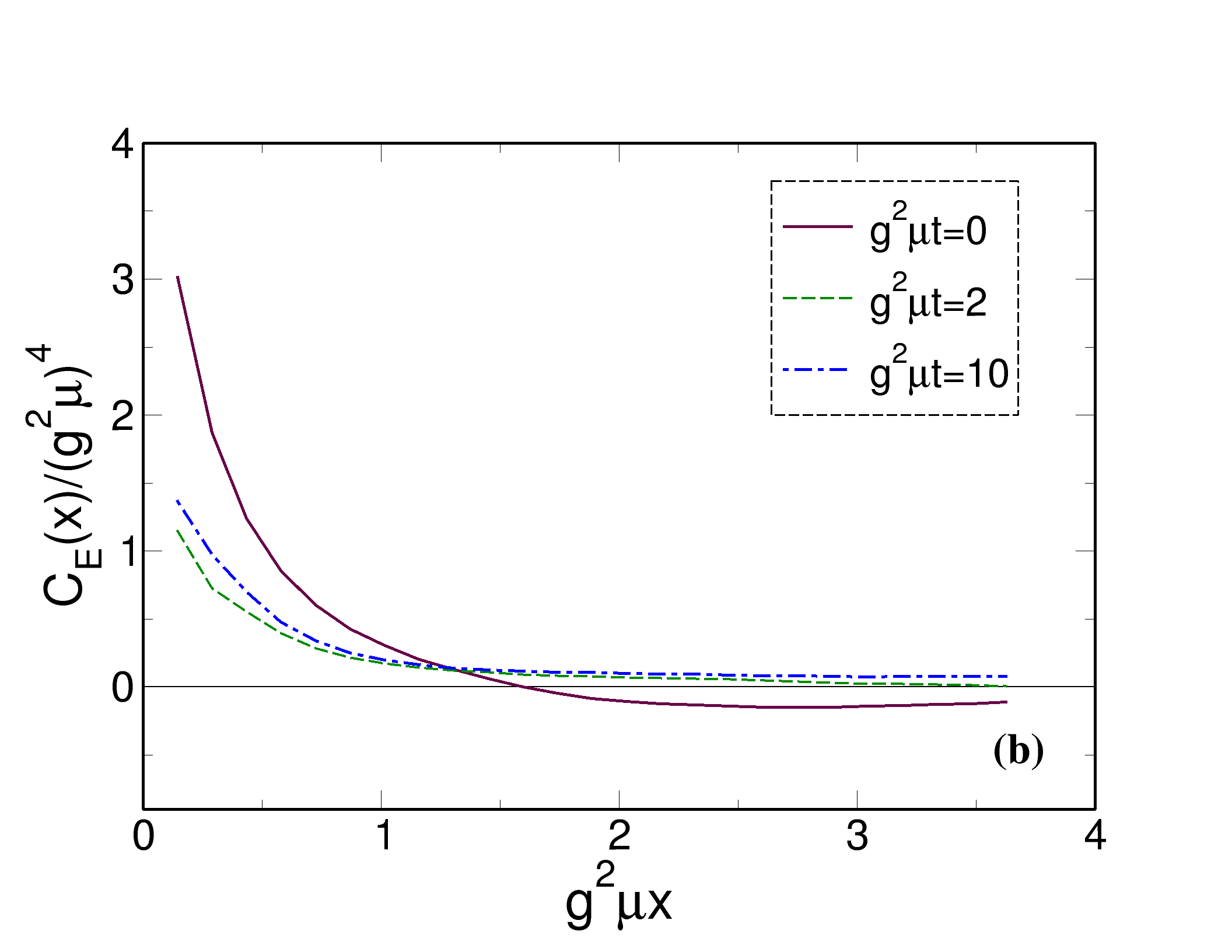}
\end{center}
\caption{\label{Fig:3a}Color-magnetic (upper panel) and color-electric (lower panel) correlators
as a function of the transverse plane coordinate, for several values of time.
Maroon solid lines corresponds to the correlators at initial time; green dashed lines denote that
at $g^2\mu t = 2$ and blue dot-dashed lines correspond to $g^2\mu t = 10$.
Lattice specifications and physical parameters are the same used for the data in Fig.~\ref{Fig:1}.}
\end{figure}

In Fig.~\ref{Fig:3a} we plot the  magnetic (upper panel) and electric (lower panel) correlation functions
versus the transverse plane coordinate, for several values of time.
Maroon solid lines corresponds to the correlators at initial time; green dashed lines denote that
at $g^2\mu t = 2$ and blue dot-dashed lines correspond to $g^2\mu t = 10$.
Lattice specifications and physical parameters are the same used for the data in Fig.~\ref{Fig:1}.
We have checked that changing $g^2\mu$ does not change qualitatively 
the results. 

We begin by discussing the result at the initial time: in fact the data
for this correlator show a couple of interesting features. First of all we notice that the correlator
decays at short distances, becoming negative for $g^2\mu x\approx 0.4$. 
Following \cite{Dumitru:2013wca} we fit this first portion
of the correlator by the standard two-dimensional screened propagator,
\begin{equation}
C_M(r) = \frac{A}{(m r)^{1/2}}e^{-m r},~~r=g^2\mu x,
\label{eq:2Dp}
\end{equation}
in order to estimate the value of the screening mass $m$ in units of $g^2\mu$: we find $m\approx 8.3$ that implies
a screening length for the magnetic modes, $\lambda_M$, equal to $g^2\mu\lambda_M\approx 0.11$.
This result shows the existence of very tiny correlation domains in the transverse plane for the Glasma, 
with a transverse size much smaller than the characteristic length scale of the system given by 
$1/g^2\mu$. The existence of this screening of the magnetic modes 
is in agreement with the one found in \cite{Dumitru:2013wca} 
where it has been related to the existence of sources and sinks of the magnetic field lines in the transverse plane.

The magnetic correlator at initial time becomes negative on a length scale, 
$\lambda_A$, approximately given by $\lambda_A\approx  0.4/g^2\mu$: this result, in qualitative agreement with
the one found in \cite{Dumitru:2014nka}, shows an anticorrelation in the initial Glasma 
that developes at $\lambda_A$, meaning that walking on the transverse plane over distances
of the order of  $\lambda_A$ it is possible to cross domains in which the magnetic field flips its sign.

On the other hand at later times the situation changes both qualitatively and quantitatively.
As a matter of fact, already for $g^2\mu t=2$ the anticorrelation disappears:
within a very short time range the CYM interaction has been capable to align the fields within domains with transverse area
$(g^2\mu)^2 A_T \approx 1$. Moreover, the fit with the two-dimensional propagator in Eq.~(\ref{eq:2Dp})
gives $g^2\mu\lambda_M\approx 0.22$ showing that the CYM interactions have enlarged 
(in fact, doubled) the correlation domains
of the magnetic fields, albeit
these correlations domains are still microscopic since their spatial extension in the transverse plane
is quite smaller than the typical size of a nucleon.

For the color-electric correlator shown in the lower panel of Fig.~\ref{Fig:3a}  we can perform a similar analysis:
in this case we find that the initial correlation length is $g^2\mu\lambda_E\approx 0.82$;
the qualitative behavior of the electric correlator agrees with that of the magnetic one,
in fact we find that anticorrelation develops for $g^2\mu\lambda_A\approx  1$ at the initial time,
and this anticorrelation disappears already for $g^2\mu t=2$. The electric correlation length does not seem to be
very affected by the CYM evolution, indeed at $g^2\mu t=3$ we find $g^2\mu\lambda_E\approx 0.76$ which,
within numerical uncertainties, agrees with the value of the correlation length at $t=0$.

\begin{figure}[t!]
\begin{center}
\includegraphics[scale=0.3]{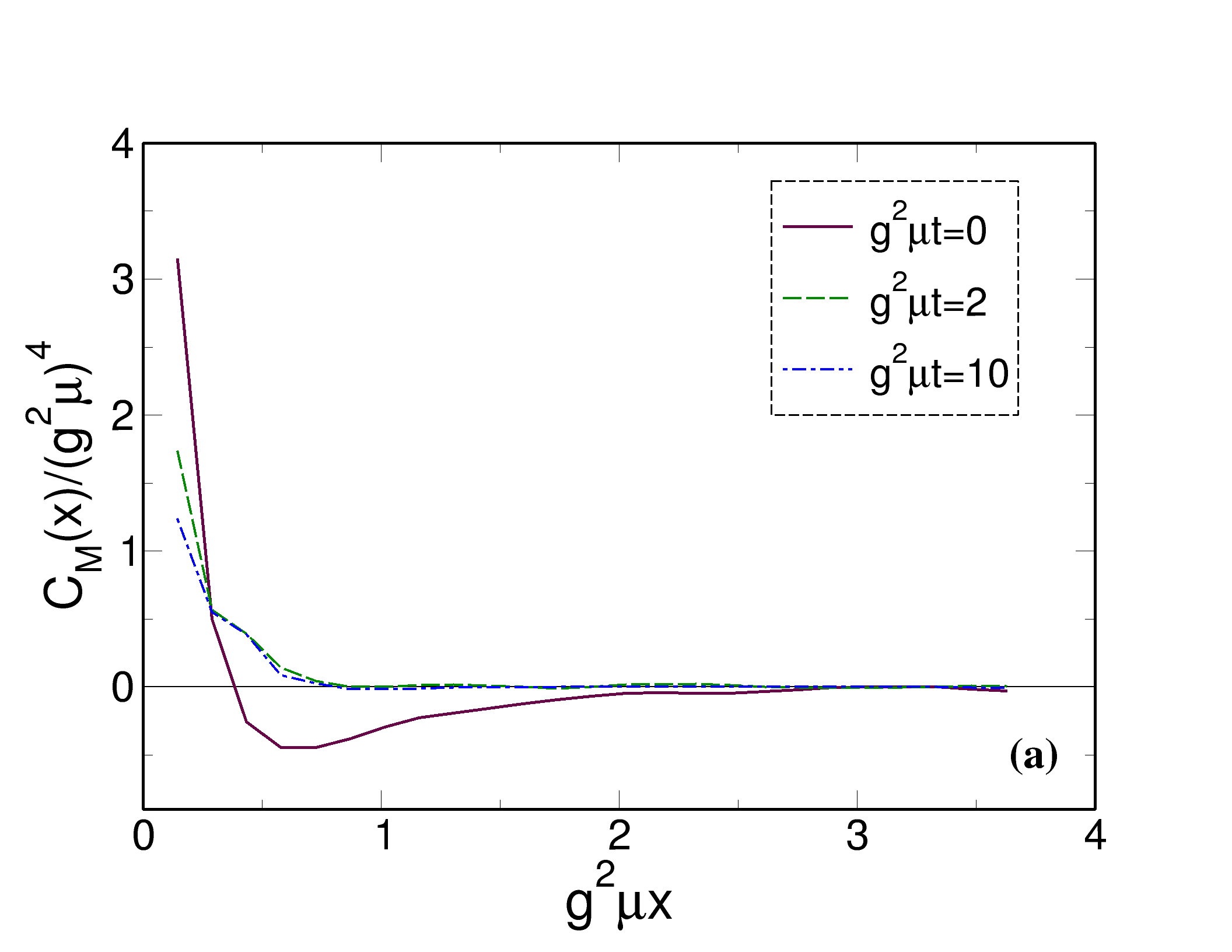}\\
\includegraphics[scale=0.3]{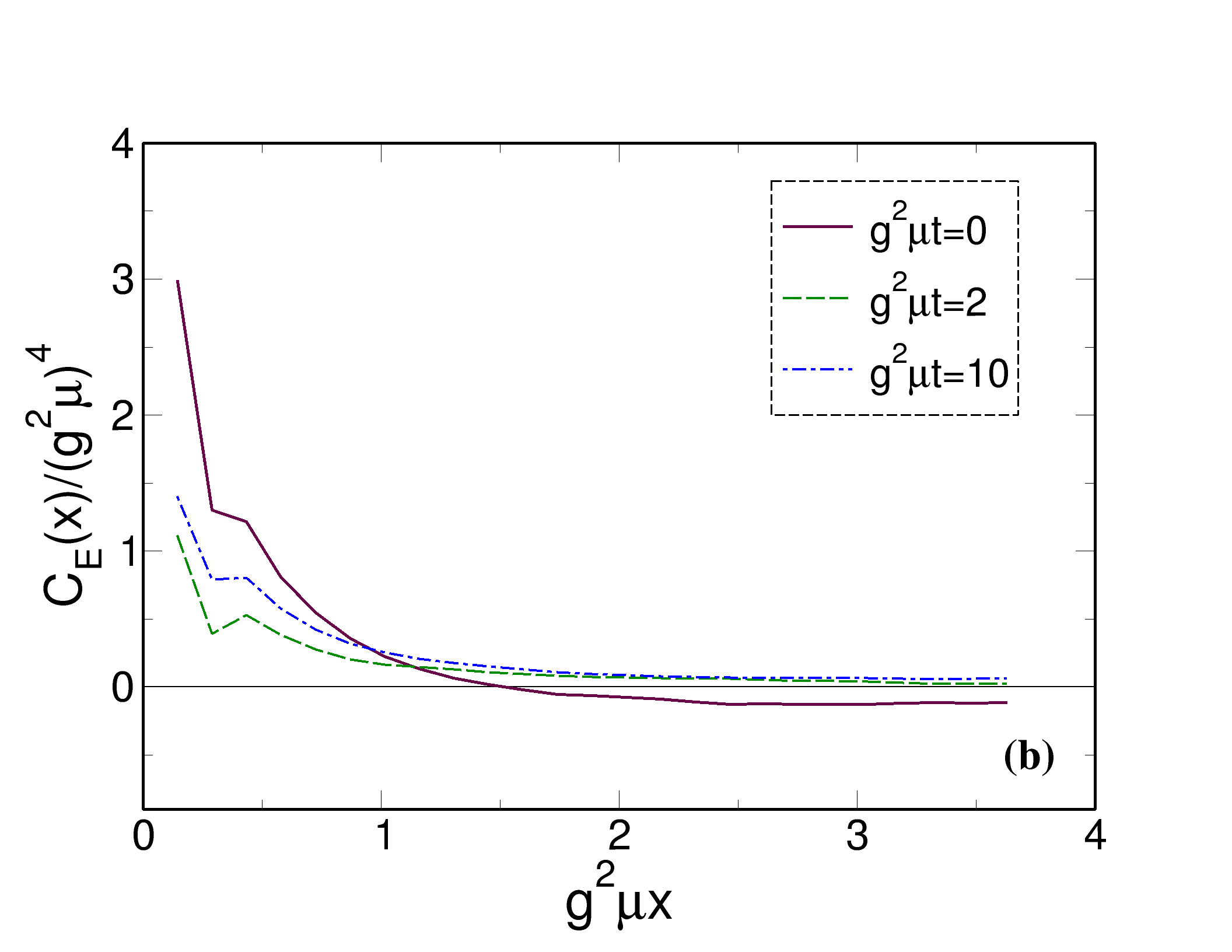}
\end{center}
\caption{\label{Fig:3aa}Color-magnetic (upper panel) and color-electric (lower panel) correlators
as a function of the transverse plane coordinate, for several values of time and for a fluctuation seed
$\Delta^2 = 2.2\times 10^{-4} g^2\mu a$ that corresponds to that used in Fig.~\ref{Fig:1}.
Maroon solid lines corresponds to the correlators at initial time; green dashed lines denote that
at $g^2\mu t = 2$ and blue dot-dashed lines correspond to $g^2\mu t = 10$.
Lattice specifications and physical parameters are the same used for the data in Fig.~\ref{Fig:1}.}
\end{figure}

For completeness, in Fig.~\ref{Fig:3aa} we plot the color-magnetic (upper panel) and color-electric (lower panel) correlators
as a function of the transverse plane coordinate, for several values of time,
in the case of the Glasma plus gaussian fluctuations with the seed 
$\Delta^2 = 2.2\times 10^{-4} g^2\mu a$ that corresponds to that used in Fig.~\ref{Fig:1}.
Maroon solid lines corresponds to the correlators at initial time; green dashed lines denote that
at $g^2\mu t = 2$ and blue dot-dashed lines correspond to $g^2\mu t = 10$.
Lattice specifications and physical parameters are the same used for the data in Fig.~\ref{Fig:1}.
We remind that we study a three dimensional box and that the correlators are computed on the transverse
plane, averaging over the volume of the box, therefore the results shown in Fig.~\ref{Fig:3aa} are obtained
averaging over the contribution of all the transverse planes corresponding to different values of 
the longitudinal coordinate. 
We find that both the magnetic and the electric correlation functions are not affected 
in a substantial way by the gaussian fluctuations; we measure a small bump of the average of the electric correlator
at small $g^2\mu x$, but taking into account the dispersion of the data the behavior of the correlator with fluctuations
is in agreement with that without fluctuations. 

\subsection{Correlation functions along the longitudinal direction}

\begin{figure}[t!]
\begin{center}
\includegraphics[scale=0.3]{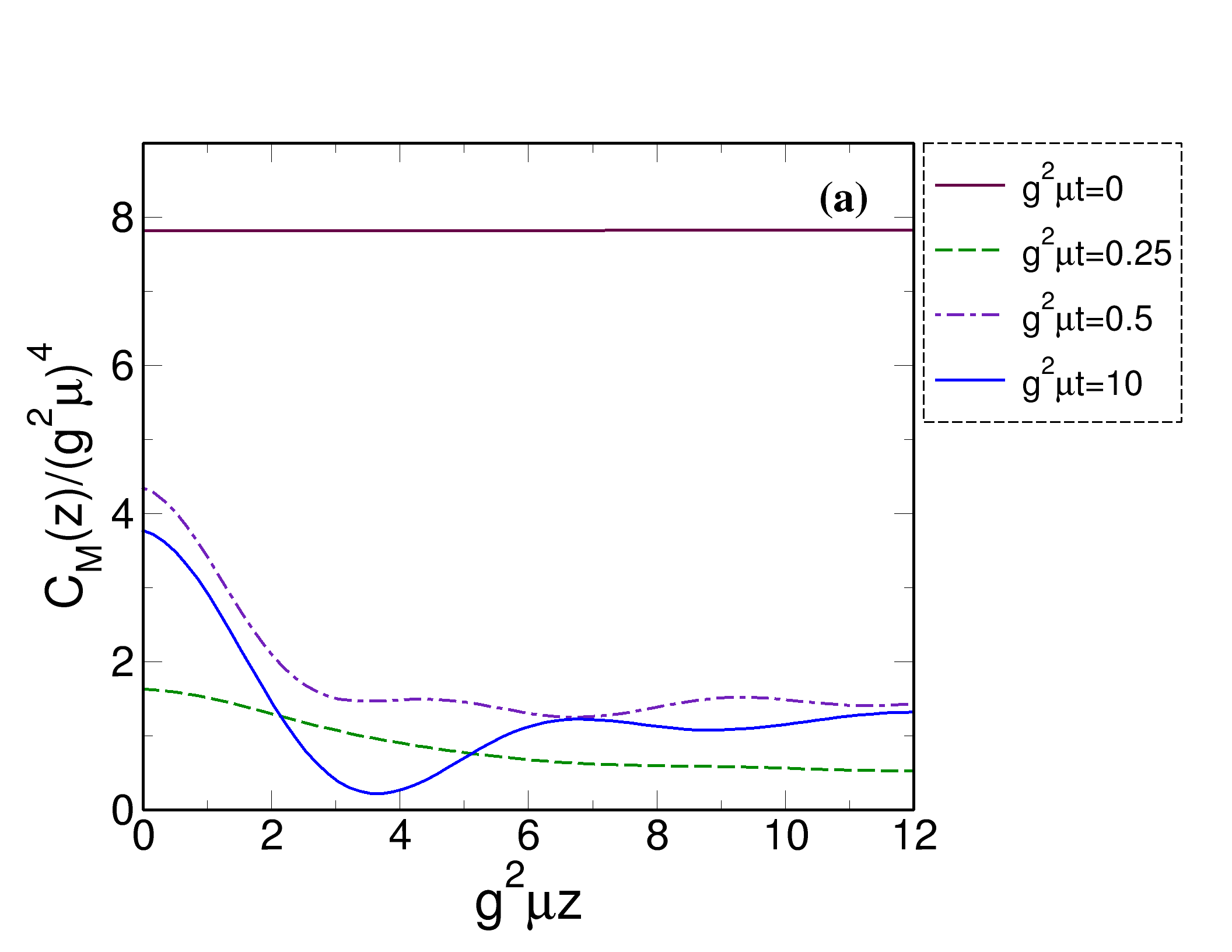}\\
\includegraphics[scale=0.3]{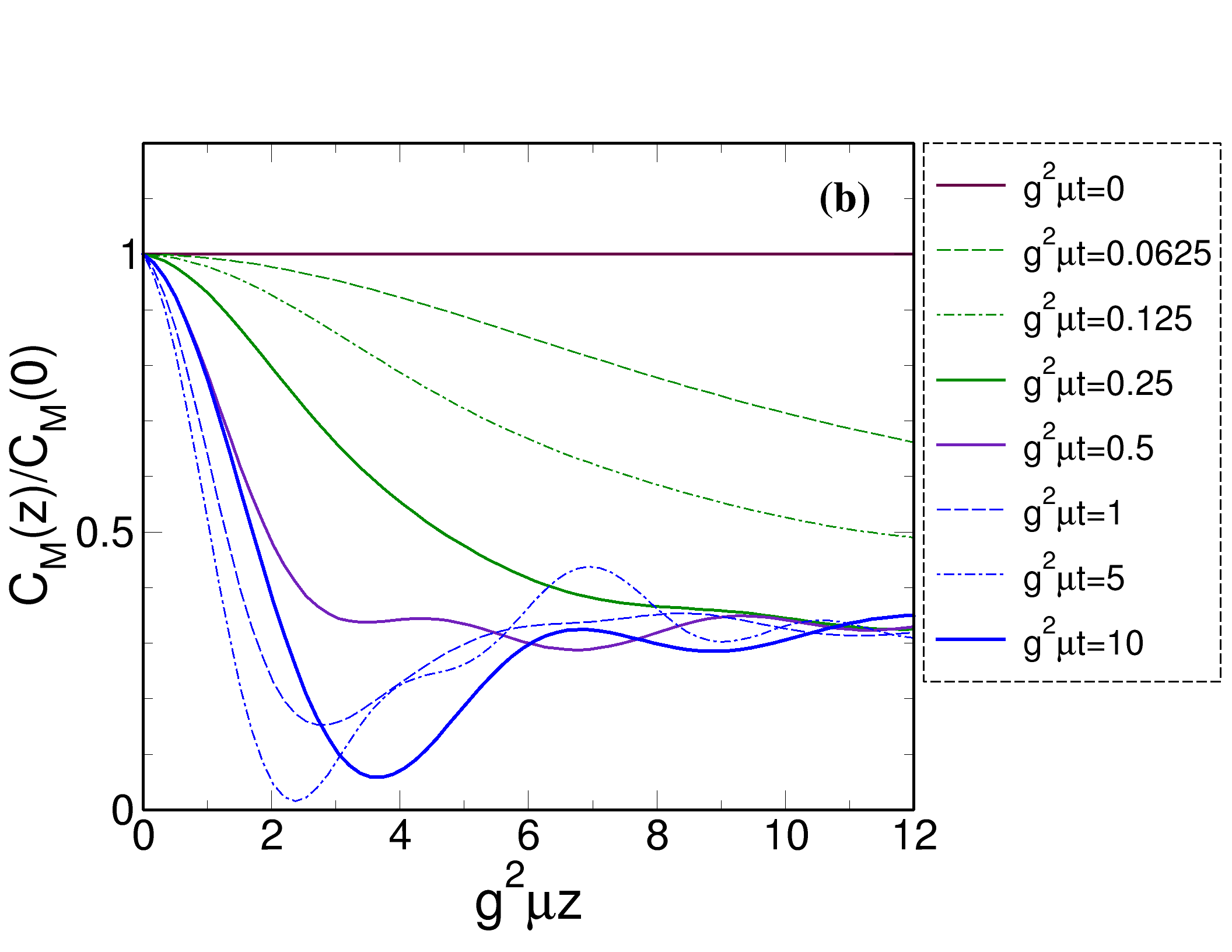}
\end{center}
\caption{\label{Fig:3aaab}Gauge invariant color-magnetic  correlators
as a function of the longitudinal coordinate. 
Panel {\bf (a)} shows the correlation function for several selected values of time: maroon line corresponds to
$t=0$, green dashed line to $g^2\mu t=0.25$, indigo dot-dashed line to $g^2\mu t=0.5$, 
and blue solid line to $g^2\mu t=10$.
Panel {\bf (b)} shows the ratio $C_M(z)/C_M(0)$ for a larger set of times.
Data shown for $g^2\mu = 1$ GeV, $g^2\mu a = 0.14$ as in Fig.~\ref{Fig:1} and
for a lattice $71\times71\times151$.}
\end{figure}

\begin{figure}[t!]
\begin{center}
\includegraphics[scale=0.3]{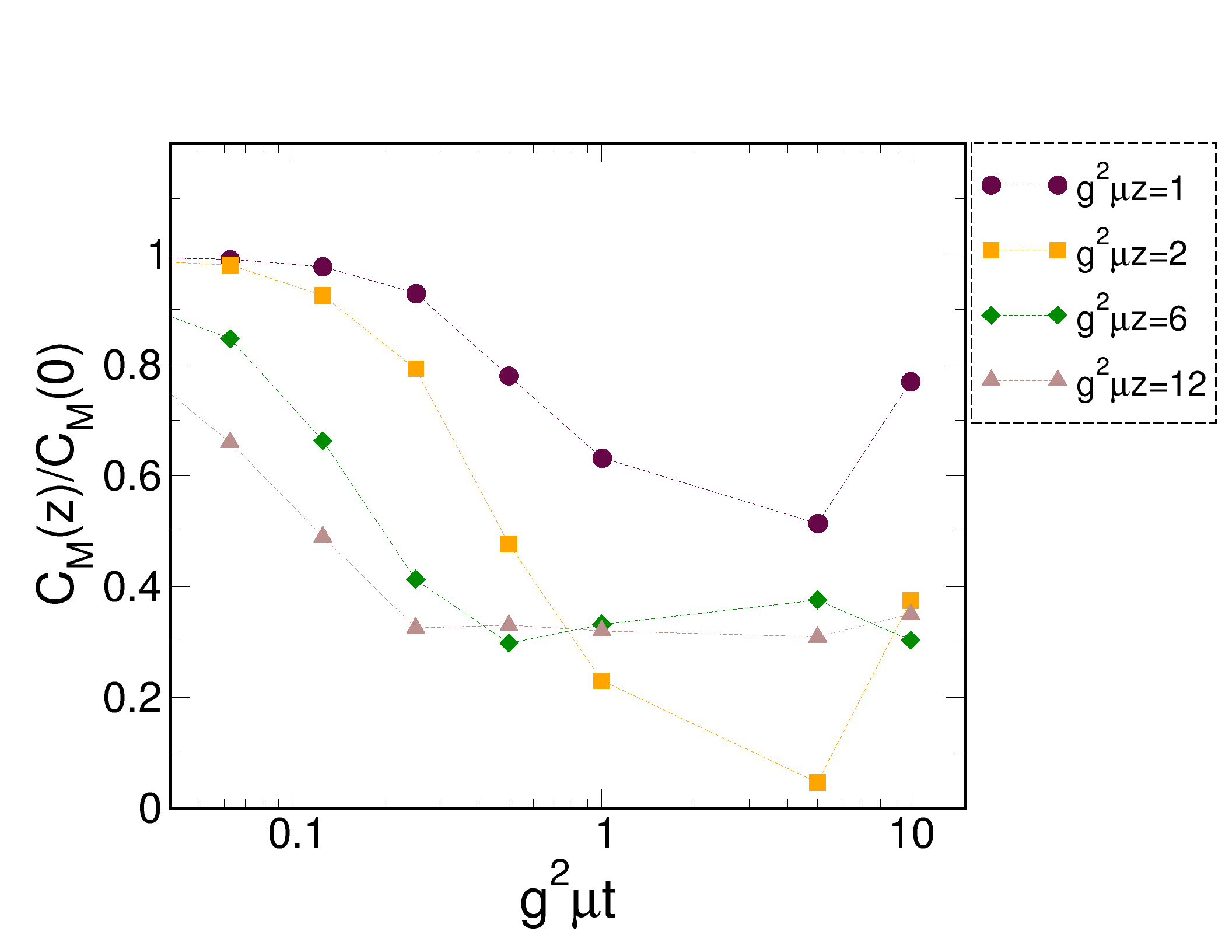}
\end{center}
\caption{\label{Fig:3aaaT}Gauge invariant color-magnetic  correlator along the longitudinal direction,
normalized to its value at $z=0$,
as a function of $g^2\mu t$ for several values of the longitudinal separation $z$.
Data have been obtained by those of panel  {\bf (b)} of Fig.~\ref{Fig:3aaab}.
Data shown for $g^2\mu = 1$ GeV, $g^2\mu a = 0.14$  and
for a lattice $71\times71\times151$.}
\end{figure}

In Fig.~\ref{Fig:3aaab} we plot the magnetic correlator as a function of the longitudinal separation $z$.
The results are shown for the case without fluctuations.
In panel {\bf (a)} we plot the correlation function for selected values of time: maroon solid line corresponds to
$t=0$, green dashed line to $g^2\mu t=0.25$, indigo dot-dashed line to $g^2\mu t=0.5$ and blue solid line to $g^2\mu t=10$.
On panel {\bf (b)} we plot the ratio $C_M(z)/C_M(0)$ for a larger set of times.
Data shown for $g^2\mu = 1$ GeV, $g^2\mu a = 0.14$ as in Fig.~\ref{Fig:1} and
for a lattice $71\times71\times151$. WIth respect to the lattice used in the previous sections we have built up 
a lattice that is longer along the longitudinal direction in order to study better these correlation functions.
In Fig.~\ref{Fig:3aaaT} we arrange data shown in panel {\bf (b)} of  Fig.~\ref{Fig:3aaab}
in the form of a normalized correlator as a function of time for several values of the longitudinal separation.
We show the normalized ratio $C_M(z)/C_M(0)$ because this quantity allows us to describe more 
clearly the behavior of the correlations along the longitudinal direction. 
As a matter of fact, as it is evident from the data in panel {\bf (a)} of Fig.~\ref{Fig:3aaab},
at each time the value of $C_M(z)$ at the origin fluctuates in particular in the very early stage,
because that corresponds to twice the average of the magnetic longitudinal energy which changes with time,
see Fig.~\ref{Fig:1}: therefore, what is really important is the $z-$variation of the correlation function with respect
to the value at $z=0$ rather than its magnitude. 

The data shown in Figg.~\ref{Fig:3aaab} and~\ref{Fig:3aaaT} are interesting for several reasons.
Firstly, we notice that although the magnetic field is $z-$independent due to the lack of longitudinal fluctuations,
the correlation function at $t>0$ is $z-$dependent. This might sound counterintuitive: 
in fact, $B_z^a$ is $z-$independent due to the lack of fluctuations so one naively would expect
the correlation function to be $z-$independent as well.
However,
we should remind that we compute the gauge-invariant correlator, $C_M$, which differs from the naive one, 
$C_\mathrm{naive}=\langle B_z^a(0) B_z^a(z)\rangle$, that would be $z-$independent; the two would 
coincide only if $A_z = 0$ which in turn happens at $t=0$ and explains why at the initial time
$C_M$ does not depend on the longitudinal coordinate.

The difference among the two correlators is clearly given by the presence of the parallel transporter along the $z$ direction,
$U=P e^{i\int_0^z dz^\prime A_z(z^\prime)}$. 
As a matter of fact, in the case without fluctuations both $B_z$ and $A_z$ are $z-$independent so 
$[A_z(z_1),A_z(z_2)]=0$ and path ordering gives the standard
exponential matrix of the integral over the full path so we can write,
for a straight path from $0$ to $z$ along the $z$ direction
\begin{equation}
U = Pe^{i\int_0^z dz^\prime A_z(z^\prime)} = e^{i\int_0^z dz^\prime A_z^a(z^\prime)\tau_a}
=e^{i z A_z^a \tau_a},
\end{equation}
with $\tau_a$ denoting the standard Pauli matrices in color space; then for each point in the transverse plane
we are left with
\begin{equation}
C_M (z)= B_a(0) B_b(z) G^{ab}(z)
\end{equation}
where
\begin{equation}
G^{ab}(z) = \mathrm{Tr}\left[\tau^a e^{i\int_0^z dz^\prime A_z^c(z^\prime)\tau_c} \tau^b 
e^{-i\int_0^z dz^\prime A_z^d(z^\prime)\tau_d}\right];
\end{equation}
noticing that $B_z$ does not depend on $z$ the term $B_a(0) B_b(z)$ is symmetric for the exchange of $a$ and $b$
and it suffices to compute the symmetric part of $G^{ab}(z)$;
this is a straightforward calculation that can be performed by using the generalized Euler idendity for $SU(2)$ matrices and it leads to 
\begin{eqnarray}
\mathrm{Sym}\left[G^{ab}(z)\right] &=& \frac{A_a^2 +(A_c^2 + A_d^2)\cos(2|Az|)}{|A|^2},~~a=b,\nonumber\\
&&
\end{eqnarray}
where $a\neq c \neq d$, and
\begin{eqnarray} 
\mathrm{Sym}\left[G^{ab}(z)\right] &=& \frac{2 A_a A_b\sin(2|Az|)^2}{|A|^2},~~a \neq b.
\end{eqnarray}
We notice the explicit $z-$dependence of the matrix elements that can not disappear even after
assuming a uniform random distribution of $A_z$ over the transverse plane.
Also notice that in the limit $A_z\rightarrow 0$ then $\mathrm{Sym}\left[G^{ab}(z)\right]  \rightarrow 1$
and the $z-$dependence disappears: the correlator in this limiting case coincides with the naive one. 
As a further check of the numerical code we have verified that if we replace the
parallel transporter with the unit matrix we obtain a $z-$independent correlation function at any time.

This result can be summarized in other words by saying that the statement that a $z-$indpedentent color-magnetic field gives a $z-$independent
correlator is not gauge invariant unless $A_z = 0$; if this was the case then the naive correlator would be gauge invariant
and independent on the longitudinal separation.
From the physical point of view, we interpret this result as the effect of the interaction of two gluon fields
separated by $z=Z$ with the background with $A_z \neq 0$:  this interaction screens the zero mode
and causes the lowering of the correlator for a large value of $g^2\mu Z$.

We notice that for small values of $g^2\mu t$ the variation of the correlator at small $z$
is slower than the one we measure for larger separation: for example, compare the maroon dots
with the brown triangles up to $g^2\mu t = 0.3$ in  Fig.~\ref{Fig:3aaaT} (we easily get the same
message from panel {\bf{(b)}} in Figg.~\ref{Fig:3aaab}). On the other hand, for
$g^2\mu t\approx 0.3$ the correlations at large separation slow down their variation and approach an approximantely
constant value for a long time range, while correlations at small separation still evolve and decrease with time.
This evolution seems to continue up to $g^2\mu t=10$. 

The picture that we read from the above results is that the classical gluon dynamics is capable to affect the long range
gauge invariant correlations of the evolving Glasma, but only within a very short time range, 
while it affects the short range dynamics for a somehow larger
time range. During the very early stage the Yang-Mills evolution lowers the long range correlations
but it is not able to cancel them completely, leaving a residual long range order: most likely this will affect the
particle production from the classical gluon field and it might be the reason of the ridge measured in proton-proton collisions.

We call the process described above as the partial gauge-invariant string breaking in the Glasma.
In fact, at the initial time the Glasma is made of $z-$independent color-magnetic fields that develop
along the longitudinal direction, almost uncorrelated in the transverse plane, so it looks like an ensemble
of color strings that extend among the two nuclei: this system is characterized by a perfect
long range order along the longitudinal direction. As the time goes on $A_z$ is formed dynamically
and this results in a quick and partial loss of the gauge-invariant correlation of the longitudinal magnetic fields with $z$;
this loss of correlation happens quickly for large separations while it is slower for smaller separations.
Eventually this results in a residual long range order, with a correlation that is almost constant for large $z$
but being approximately only the $30\%$ of the correlation measured for small $z$.
For this reason we say that the original Glasma strings have been partly broken in a gauge invariant way by the dynamics. 
We have verified that this evolution of the longitudinal correlator happens also for the electric correlation function,
and that changing the longitudinal size of the box does not change qualitatively the results.

\begin{figure}[t!]
\begin{center}
\includegraphics[scale=0.3]{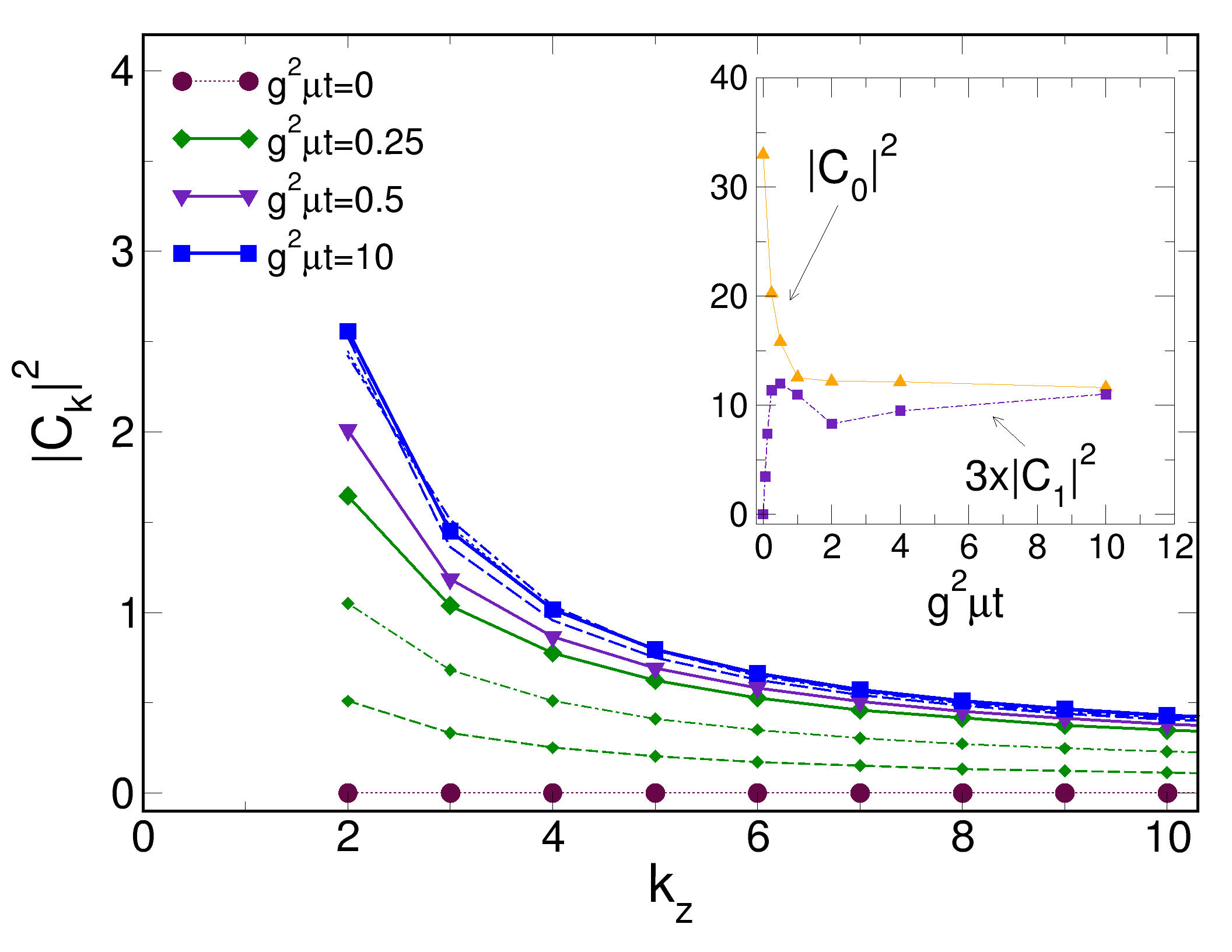}
\end{center}
\caption{\label{Fig:3png}Longitudinal Fourier transform of the correlation function at different times. 
Lines and colors conventions agree with those used in panel {\bf (b)} of Fig.~\ref{Fig:3aaab}.
Data shown for $g^2\mu = 1$ GeV, $g^2\mu a = 0.14$  and
for a lattice $71\times71\times151$.
In the inset we show $|C_0|^2$ and $|C_1|^2$ versus time.}
\end{figure}

The gauge invariant partial loss of correlation can be described also in the longitudinal momentum space.
In Fig.~\ref{Fig:3png} we plot the squared magnitude of the Fourier coefficients of the correlation function,
the latter being defined as
\begin{equation}
C_{k} = \sum_{{n}=0}^{N_z} \frac{C_M(n)}{C_M(0)}e^{i\frac{2\pi k n}{N_z -1}},
\end{equation}
where $N_z = 150$ corresponds to the longitudinal lattice size and $n$ is
the discrete index representing a lattice site in the longitudinal direction. In Fig.~\ref{Fig:3png}
the main panel represents $|C_k|^2$ for $k\geq 2$ at different times, while for the sake of keeping the picture clear
we have collected
the results for $|C_0|^2$ and $|C_1|^2$ in the inset of the same figure.
From these results we notice that at the initial time the only contribution to the correlator comes from the zero mode,
which is obvious because at this time the field has no $z-$dependence.
On the other hand, the dynamics lowers the zero mode coefficient very quickly; in fact,
for $g^2\mu t \leq 0.5$ we find that $|C_0|$ changes approximately of a factor of $2$
then it stays almost constant and nonzero.
At the same time modes with $k\neq 0$ are populated and these are responsible of the acquired $z-$dependence of the
correlation function.
We notice that while the zero-mode
as well as the low momentum modes evolve mainly for  $g^2\mu t \leqslant 0.3 $,
the higher modes have a relative substantial variation also for larger times.
In the future we will perform simulations with larger lattices in order to have a better resolution of the Fourier
transform of this correlation function.

\begin{figure}[t!]
\begin{center}
\includegraphics[scale=0.3]{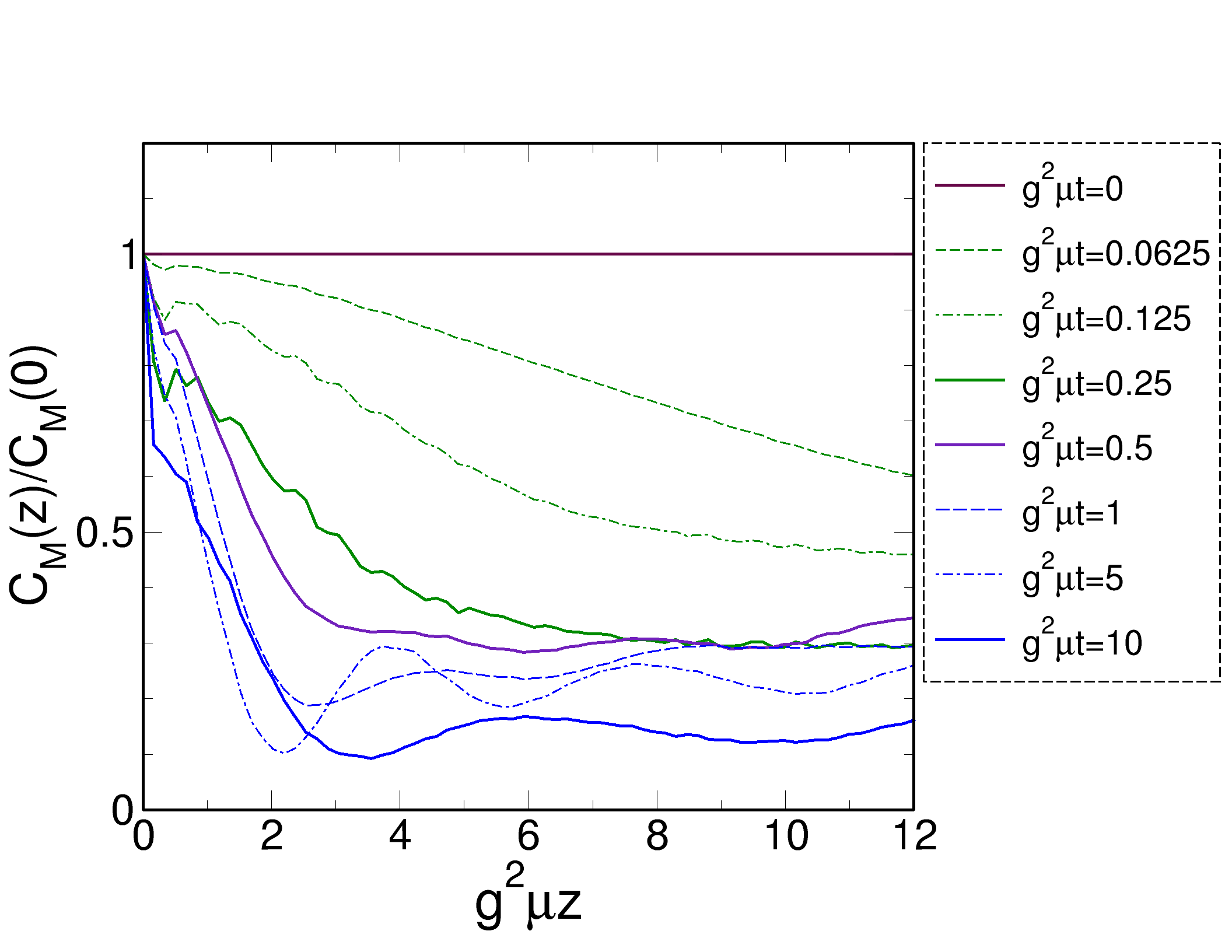}
\end{center}
\caption{\label{Fig:3aaa}Color-magnetic  correlators
as a function of the longitudinal coordinate, for several values of time and for a fluctuation seed
$\Delta^2 = 2.2\times 10^{-4} g^2\mu a$.  Colors and line conventions agree with panel (b) of Fig.~\ref{Fig:3aaab}.
Data shown for $g^2\mu = 1$ GeV, $g^2\mu a = 0.14$  and
for a lattice $71\times71\times151$.}
\end{figure}

In Fig.~\ref{Fig:3aaa} we plot the correlation function along the longitudinal direction, normalized to the value at $z=0$,
in the case fluctuations are added on the top of the Glasma with seed
$\Delta^2 = 2.2\times 10^{-4} g^2\mu a$. We find that the correlations of the magnetic field
are affected by the fluctuations. In particular, in the very early stage the main effect of the fluctuations is the damping of the oscillations
that we have found for the case without fluctuations, compare Fig.~\ref{Fig:3aaab} with Fig.~\ref{Fig:3aaa}.
For larger $z$  we do not observe the freezing of the correlations that we have
found without fluctuations; in fact comparing the data at $g^2\mu t=10$ in the two cases we notice that
in the case of fluctuations the correlator at large $z$ decreases to $\approx 15\%$ of the value at $z=0$,
while in the case without fluctuations the correlator decreases to $\approx 30\%$ of the value at $z=0$. 
This suggests that fluctuations enhance the breaking of the color strings.

\begin{figure}[t!]
\begin{center}
\includegraphics[scale=0.3]{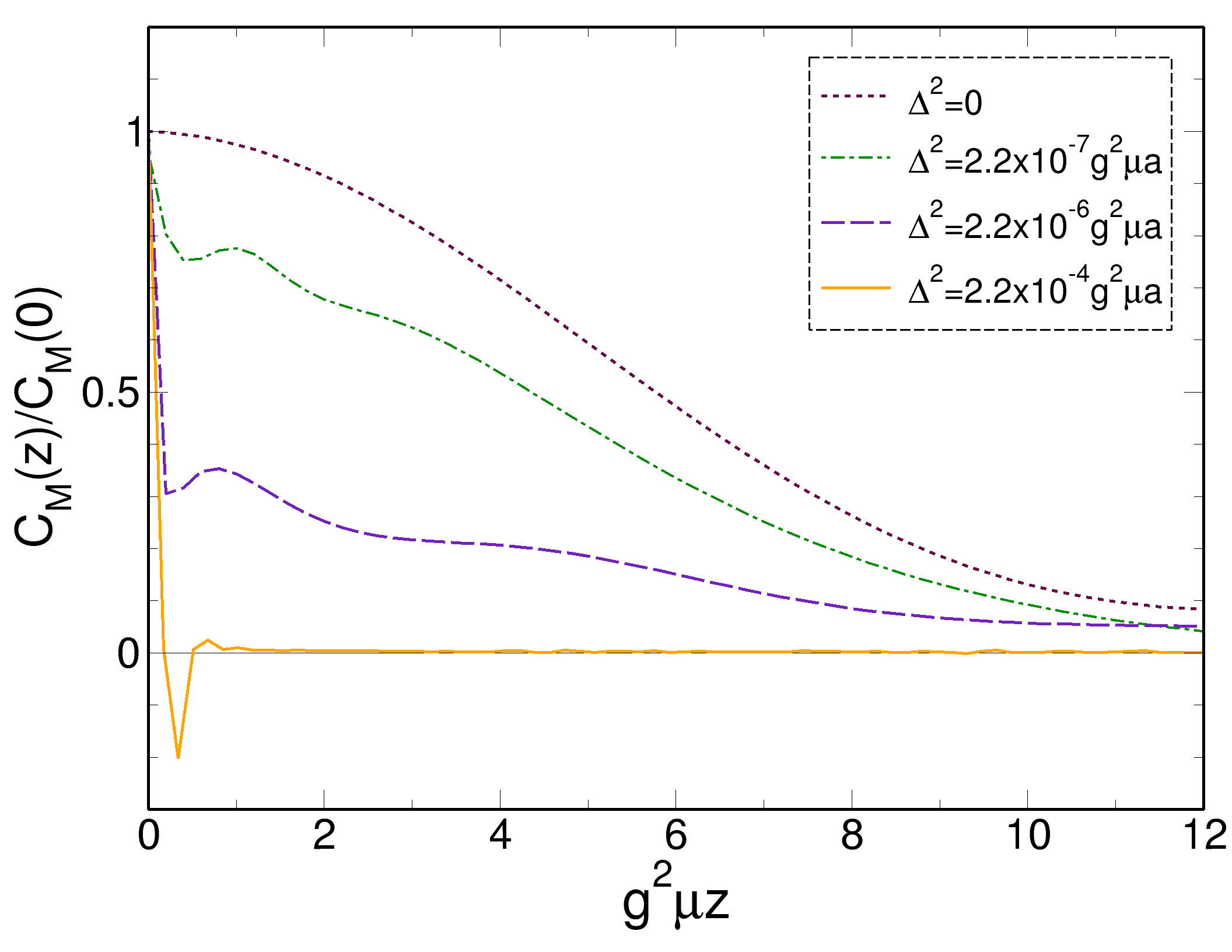}
\end{center}
\caption{\label{Fig:3aaaBBB}Correlation function of the color-magnetic field along the longitudinal direction at $g^2\mu t = 50$
for three values of the fluctuation seed: dot data stand for the case $\Delta=0$, dot-dashed green line corresponds to 
$\Delta^2 = 2.2\times 10^{-7} g^2\mu a$, dashed indigo line stands for $\Delta^2 = 2.2\times 10^{-6} g^2\mu a$,
finally solid orange line denotes the case $\Delta^2 = 2.2\times 10^{-4} g^2\mu a$. }
\end{figure}

In order to put a firm statement on the enhancement of string breaking by fluctuations
we have followed the evolution of the system up to a larger time for different values of the fluctuation seed.
The results are collected in Fig.~\ref{Fig:3aaaBBB} where we plot the correlation of the
color-magnetic field along the longitudinal direction at $g^2\mu t = 50$
for three values of the fluctuation seed: dotted maroon data stand for the case without fluctuations,
dot-dashed green line corresponds to 
$\Delta^2 = 2.2\times 10^{-7} g^2\mu a$, dashed indigo line stands for $\Delta^2 = 2.2\times 10^{-6} g^2\mu a$,
finally solid orange line denotes the case $\Delta^2 = 2.2\times 10^{-4} g^2\mu a$. 
The results in Fig.~\ref{Fig:3aaaBBB}  show that 
at a given large time fluctuation tend to lower the magnitude of the correlator;
in fact in the case $\Delta^2 = 2.2\times 10^{-4} g^2\mu a$ the correlation seems to vanish already for 
$g^2\mu z \approx 1$ whole being clearly nonvanishing for smaller values of $\Delta$ as well as for $\Delta=0$. 
We can then conclude that fluctuations accelerate the  breaking of the Glasma strings.
We remark however that $g^2\mu t =50$ corresponds to the physical time $t \approx 10$ fm/c which is quite
beyond the expected range of validity of the CYM approach that instead should describe the system
up to $t\approx 1$ fm/c; therefore, in realistic  collisions
the results in Figg.~\ref{Fig:3aaab}  and~\ref{Fig:3aaa} suffice and we expect only a partial breaking of the Glasma strings
to occur in these processes.

Concluding this section, we have found that the CYM dynamics leads to a partial loss of the gauge invariant correlation along the longitudinal
direction (the naive correlator would  be $z-$independent) even without fluctuations, and that fluctuations
enhance the string breaking since they accelerate the degradation of the gauge invariant correlation function.
We have shown the results for the color-magnetic strings: for the electric fields we have obtained similar results.
It is useful to remark that these results have been obtained in simulations of the color fields in a static box: 
the effect of the longitudinal expansion on the correlations will be the subject of a forthcoming article.

\section{Conclusions and outlook}

We have studied the evolution of the color fields produced in high energy nuclear
collisions, namely the Glasma with gaussian fluctuations, 
by means of classical Yang-Mills equations that we have solved numerically on a lattice.  
Neglecting in this first study the longitudinal expansion, that we will consider in a near future work, 
we have followed the evolution of the pressures of the system 
and computed the effect of the fluctuations in the early stage, up to
$t\approx 2$ fm/c: 
this early stage roughly corresponds to the time range in which the classical Yang-Mills dynamics is
relevant for potential applications to high energy collisions. 

We have considered one representative value of $g^2\mu$, namely $g^2\mu=1$ GeV\footnote{We have checked that
changing the value of $g^2\mu$ does not affect the results qualitatively.}
 and following the evolution up to 
$g^2\mu t=10$, that  corresponds to 
a maximum physical time of $t\approx 2$ fm/c: this time range is 
already beyond that in which the classical field theory description can have
an interest for realistic collisions.
We have set the
transverse area to $A_T = 4$ fm$^2$.  Among other things, this choice allows for a small lattice spacing
even if the lattice size is not large. Since we solve the equations of motion
applying periodic boundary conditions, the solutions found here should represent quite
well also the evolution of the color fields in the central region of nucleus-nucleus collisions.

We have measured the ratio of the longitudinal over the
transverse pressure, $P_L/P_T$, as a function of time; this quantity is usually taken as the reference
for studying isotropization of the system, a perfect isotropy being achieved if $P_L=P_T$.
We have found that unless the fluctuations carry a substantial amount of the Glasma energy
density at the initial time, they do not change drastically the evolution of $P_L/P_T$ in the early stage.
For example, if the if the initial disturbances carry about the $65\%$ of the energy to the Glasma then
the system is capable to obtain a fair amount of isotropy, $P_L/P_T\approx0.8$ within $g^2\mu t = 1$. 

Our limited computing power does not allow us to perform long time runs, but it will be important to do these in the near
future: in fact, it is well known that
secondary instabilities trigger a rise of $P_L/P_T$ at very large times in the case of small fluctuations size \cite{Berges:2012cj},
and that the time at which these instabilities become important depends logarithmically on the fluctuation size.
Computing this in the future will be necessary to understand if the instabilities are the leading mechanism for the more efficient
isotropization in case of larger fluctuation seeds, or if a collisional-like dynamics sets in. We will address this problem in a 
forthcoming article. For the time being we take our results as preparatory of a more complete work,
emphasizing that what we have found suggests that it is possible that the classical Yang-Mills dynamics leads to a fairly isotropic state within 
a short time range, regardless of the actual mechanism used to produce the isotropy, but that in order to achieve this
the fluctuations have to be large enough.
This result might suggest that the fair amount of isotropy invoked in hydrodynamics simulations
can be the consequence of an initial state in which fluctuations are substantial;
on the other hand, we are forced to take this conclusion with a grain of salt, since 
it might put doubts on the use of the classical dynamics for the description
of the early stage of high energy nuclear collisions: indeed, if fluctuations are substantial,
and being them of a quantum nature, it might be not safe to apply a classical theory to describe the evolution
of the system.
The discussion of this complicated topic is however too far from the purpose
of our simple work, therefore we limit ourselves to summarize our results and we leave the discussion
to future works.

Finally, we have measured the 
color-magnetic and color-electric correlation functions. We have found a qualitative agreement  with previous 
studies \cite{Dumitru:2013wca,Dumitru:2014nka};
moreover, we have found that the gaussian fluctuations hardly affect the correlation functions.
We support the idea that at the initial time the Glasma fields have anticorrelation
in the transverse plane, and that the CYM evolution aligns the fields in a time range
$g^2\mu t = O(1)$ on a transverse area $(g^2\mu)^2 A_T = O(1)$.

A novelty that we have brought with our study has been that of the gauge invariant correlators along the longitudinal direction;
these are important in order to address the question of the evolution of the color strings.
We have found that also in absence of fluctuations in the initial state, the gauge invariant magnetic correlator  
decreases at large $z$; this is different from what happens to the naive, gauge dependent correlation function
which instead remains $z-$indepentent. We call this phenomenon as the gauge invariant partial string breaking.
The string breaking is related to the fact that
at the initial time the correlation function along the longitudinal direction is $z-$independent so the field looks like
a string that connects the two transverse planes, while this
correlation function becomes $z-$dependent for $t>0$ and in particular the correlation for large $z$
is smaller than the one we measure near the origin. 
It is gauge invariant since it is obtained studying a gauge invariant correlation function; it is partial
since we still find some correlation at large distance. 
We remark that this has been obtained for simulations
of the classical dynamics of color fields in a static box: most likely these correlations will experience a more
substantial decay in the expanding case that we will study in the near future.
The difference between the naive and the gauge invariant correlation functions is given formally by the
presence of the parallel transporter which in turn depends on $A_z^a$: this is zero in the initial state
but it becomes nonzero because of the Yang-Mills evolution, therefore the propagation of the gluon field
along the longitudinal direction is affected by the interaction with this $A_z$ background.
We have found that the dynamics at large $z$ is a bit faster than the one at smaller $z$:
indeed, the correlations at large $z$ evolve within $g^2\mu t \leqslant 0.3 $ then correlations seem to freeze while those at small $z$
still evolve in the full time range studied. We have confirmed this by the computation
of the Fourier transform along the longitudinal direction, finding that while the zero-mode
as well as the low momentum modes evolve drastically for  $g^2\mu t \leqslant 0.3 $,
the higher modes have a relative substantial but slow variation also for larger times.
In the future we will perform simulations with larger lattices in order to have a better resolution of the Fourier
transform of this correlation function.

Finally, we have computed the effect of the fluctuations
on the correlations along the longitudinal direction. We have found that 
fluctuations enhance this gauge invarant string breaking since they accelerate the decay of the longitudinal
correlation function, see Figg.~\ref{Fig:3aaa} and~\ref{Fig:3aaaBBB}.
However, we have found a substantial breaking of the color strings only for times
that are well beyond the alleged applicability of the CYM approach to realistic collisions;
it is likely that for these processes the results in Figg.~\ref{Fig:3aaab} and~\ref{Fig:3aaa} suffice to describe the correlations
of the gluon fields in the early stage, thus that substantial correlation remains for large longitudinal separation
and the Glasma strings are only partially broken even when fluctuations are included.

We think that the most important improvement of the work presented here is the 
introduction of the longitudinal expansion
in order to describe more closely the early stage of the system created by the collision. 
In addition to this, we also would like to improve our simulation code by means of a parallelizzation
that will allow us to perform the long time runs like those in \cite{Romatschke:2005pm,Fukushima:2013dma,Berges:2012cj}
and study the interplay of the collisional dynamics with the Glasma instabilities.
We plan to report on these topics in future works.

\begin{acknowledgements}
M. R. is indebted to F. Gelis and Y. Nara for the many suggestions given
during the preparation of this article. 
The work of M. R. has been supported by the National Natural Science Foundation of China 
(11575190 and 11475110) and the Chinese Academy of Sciences President International Fellowship 
Initiative (2015PM008).
G.X.P and V.G would like to thank the National Natural Science Foundation of China (11575190 and 11475110) and the 
President International Fellowship Initiative (2016VMA063).
Numerical calculations have been performed on the QGPDyn Cluster of Catania University and on the
Cluster of School of Nuclear Science and Technology in Lanzhou University. 
\end{acknowledgements}


\begin{thebibliography}{99}

\bibitem{McLerran:1993ni} 
  L.~D.~McLerran and R.~Venugopalan,
  Phys.\ Rev.\ D {\bf 49}, 2233 (1994)
  [hep-ph/9309289].
\bibitem{McLerran:1993ka} 
  L.~D.~McLerran and R.~Venugopalan,
  Phys.\ Rev.\ D {\bf 49}, 3352 (1994)
  [hep-ph/9311205].
\bibitem{McLerran:1994vd} 
  L.~D.~McLerran and R.~Venugopalan,
  Phys.\ Rev.\ D {\bf 50}, 2225 (1994)
  [hep-ph/9402335].
  
\bibitem{Gelis:2010nm} 
  F.~Gelis, E.~Iancu, J.~Jalilian-Marian and R.~Venugopalan,
  Ann.\ Rev.\ Nucl.\ Part.\ Sci.\  {\bf 60}, 463 (2010).
\bibitem{Iancu:2003xm} 
  E.~Iancu and R.~Venugopalan,
  In *Hwa, R.C. (ed.) et al.: Quark gluon plasma* 249-3363.
\bibitem{McLerran:2008es} 
  L.~McLerran,
  arXiv:0812.4989 [hep-ph];
  hep-ph/0402137.
  
\bibitem{Gelis:2012ri} 
  F.~Gelis,
  Int.\ J.\ Mod.\ Phys.\ A {\bf 28}, 1330001 (2013).

\bibitem{Kovner:1995ja} 
  A.~Kovner, L.~D.~McLerran and H.~Weigert,
  Phys.\ Rev.\ D {\bf 52}, 6231 (1995)
  doi:10.1103/PhysRevD.52.6231
  [hep-ph/9502289].

\bibitem{Kovner:1995ts} 
  A.~Kovner, L.~D.~McLerran and H.~Weigert,
  Phys.\ Rev.\ D {\bf 52}, 3809 (1995)
  doi:10.1103/PhysRevD.52.3809
  [hep-ph/9505320].

\bibitem{Gyulassy:1997vt} 
  M.~Gyulassy and L.~D.~McLerran,
  Phys.\ Rev.\ C {\bf 56}, 2219 (1997)
  doi:10.1103/PhysRevC.56.2219
 [nucl-th/9704034].


\bibitem{Lappi:2006fp} 
  T.~Lappi and L.~McLerran,
  Nucl.\ Phys.\ A {\bf 772}, 200 (2006)
  doi:10.1016/j.nuclphysa.2006.04.001
  [hep-ph/0602189].


\bibitem{Fries:2006pv} 
  R.~J.~Fries, J.~I.~Kapusta and Y.~Li,
  nucl-th/0604054.
  
\bibitem{Chen:2015wia} 
  G.~Chen, R.~J.~Fries, J.~I.~Kapusta and Y.~Li,
  Phys.\ Rev.\ C {\bf 92}, no. 6, 064912 (2015)
  doi:10.1103/PhysRevC.92.064912
  [arXiv:1507.03524 [nucl-th]].

\bibitem{Krasnitz:2000gz} 
  A.~Krasnitz and R.~Venugopalan,
  Phys.\ Rev.\ Lett.\  {\bf 86}, 1717 (2001)
  doi:10.1103/PhysRevLett.86.1717
  [hep-ph/0007108].
  
  
\bibitem{Krasnitz:2001qu} 
  A.~Krasnitz, Y.~Nara and R.~Venugopalan,
  Phys.\ Rev.\ Lett.\  {\bf 87}, 192302 (2001)
  doi:10.1103/PhysRevLett.87.192302
  [hep-ph/0108092].
  
  
\bibitem{Krasnitz:2003jw} 
  A.~Krasnitz, Y.~Nara and R.~Venugopalan,
  Nucl.\ Phys.\ A {\bf 727}, 427 (2003)
  doi:10.1016/j.nuclphysa.2003.08.004
  [hep-ph/0305112].

      
\bibitem{Fukushima:2006ax} 
  K.~Fukushima, F.~Gelis and L.~McLerran,
  Nucl.\ Phys.\ A {\bf 786}, 107 (2007)
  doi:10.1016/j.nuclphysa.2007.01.086
  [hep-ph/0610416].


\bibitem{Fujii:2008km} 
  H.~Fujii, K.~Fukushima and Y.~Hidaka,
  Phys.\ Rev.\ C {\bf 79}, 024909 (2009)
  doi:10.1103/PhysRevC.79.024909
  [arXiv:0811.0437 [hep-ph]].


\bibitem{Fukushima:2013dma} 
  K.~Fukushima,
  Phys.\ Rev.\ C {\bf 89}, no. 2, 024907 (2014)
  doi:10.1103/PhysRevC.89.024907
  [arXiv:1307.1046 [hep-ph]].



\bibitem{Romatschke:2005pm} 
  P.~Romatschke and R.~Venugopalan,
  Phys.\ Rev.\ Lett.\  {\bf 96}, 062302 (2006)
  doi:10.1103/PhysRevLett.96.062302
  [hep-ph/0510121].


\bibitem{Romatschke:2006nk} 
  P.~Romatschke and R.~Venugopalan,
  Phys.\ Rev.\ D {\bf 74}, 045011 (2006)
  doi:10.1103/PhysRevD.74.045011
  [hep-ph/0605045].


\bibitem{Fukushima:2011nq} 
  K.~Fukushima and F.~Gelis,
  Nucl.\ Phys.\ A {\bf 874}, 108 (2012.
  doi:10.1016/j.nuclphysa.2011.11.003
  [arXiv:1106.1396 [hep-ph]].

\bibitem{Iida:2014wea} 
  H.~Iida, T.~Kunihiro, A.~Ohnishi and T.~T.~Takahashi,
  arXiv:1410.7309 [hep-ph].


\bibitem{Gelis:2013rba} 
  T.~Epelbaum and F.~Gelis,
  Phys.\ Rev.\ Lett.\  {\bf 111}, 232301 (2013)
  doi:10.1103/PhysRevLett.111.232301
  [arXiv:1307.2214 [hep-ph]].

\bibitem{Epelbaum:2013waa} 
  T.~Epelbaum and F.~Gelis,
  Phys.\ Rev.\ D {\bf 88}, 085015 (2013)
  doi:10.1103/PhysRevD.88.085015
  [arXiv:1307.1765 [hep-ph]].
   
\bibitem{Ryblewski:2013eja} 
  R.~Ryblewski and W.~Florkowski,
  Phys.\ Rev.\ D {\bf 88}, 034028 (2013)
  doi:10.1103/PhysRevD.88.034028
  [arXiv:1307.0356 [hep-ph]].
 
\bibitem{Ruggieri:2015yea} 
  M.~Ruggieri, A.~Puglisi, L.~Oliva, S.~Plumari, F.~Scardina and V.~Greco,
  Phys.\ Rev.\ C {\bf 92}, 064904 (2015)
  doi:10.1103/PhysRevC.92.064904
  [arXiv:1505.08081 [hep-ph]].
 
 
\bibitem{Tanji:2011di} 
  N.~Tanji and K.~Itakura,
  Phys.\ Lett.\ B {\bf 713}, 117 (2012)
  doi:10.1016/j.physletb.2012.05.043
  [arXiv:1111.6772 [hep-ph]].


\bibitem{Berges:2012cj} 
  J.~Berges and S.~Schlichting,
  Phys.\ Rev.\ D {\bf 87}, no. 1, 014026 (2013)
  doi:10.1103/PhysRevD.87.014026
  [arXiv:1209.0817 [hep-ph]].
  
  
\bibitem{Berges:2013fga} 
  J.~Berges, K.~Boguslavski, S.~Schlichting and R.~Venugopalan,
  Phys.\ Rev.\ D {\bf 89}, no. 11, 114007 (2014)
  doi:10.1103/PhysRevD.89.114007
  [arXiv:1311.3005 [hep-ph]].
  
  
\bibitem{Berges:2013lsa} 
  J.~Berges, K.~Boguslavski, S.~Schlichting and R.~Venugopalan,
  JHEP {\bf 1405}, 054 (2014)
  doi:10.1007/JHEP05(2014)054
  [arXiv:1312.5216 [hep-ph]].

\bibitem{Berges:2013eia} 
  J.~Berges, K.~Boguslavski, S.~Schlichting and R.~Venugopalan,
  Phys.\ Rev.\ D {\bf 89}, no. 7, 074011 (2014)
  doi:10.1103/PhysRevD.89.074011
  [arXiv:1303.5650 [hep-ph]].


\bibitem{Li:2016eqr} 
  M.~Li and J.~I.~Kapusta,
  Phys.\ Rev.\ C {\bf 94}, no. 2, 024908 (2016)
  doi:10.1103/PhysRevC.94.024908
  [arXiv:1602.09060 [nucl-th]].


\bibitem{Kurkela:2015qoa} 
  A.~Kurkela and Y.~Zhu,
  Phys.\ Rev.\ Lett.\  {\bf 115}, no. 18, 182301 (2015)
  doi:10.1103/PhysRevLett.115.182301
  [arXiv:1506.06647 [hep-ph]].
  
  
\bibitem{Bellantuono:2015hxa} 
  L.~Bellantuono, P.~Colangelo, F.~De Fazio and F.~Giannuzzi,
  JHEP {\bf 1507}, 053 (2015)
  doi:10.1007/JHEP07(2015)053
  [arXiv:1503.01977 [hep-ph]].
  



\bibitem{Dumitru:2013wca} 
  A.~Dumitru, H.~Fujii and Y.~Nara,
  Phys.\ Rev.\ D {\bf 88}, 031503 (2013)
  doi:10.1103/PhysRevD.88.031503
  [arXiv:1305.2780 [hep-ph]].


\bibitem{Dumitru:2014nka} 
  A.~Dumitru, T.~Lappi and Y.~Nara,
  Phys.\ Lett.\ B {\bf 734}, 7 (2014)
  doi:10.1016/j.physletb.2014.05.005
  [arXiv:1401.4124 [hep-ph]].







  
  



\bibitem{Kovchegov:1996ty} 
  Y.~V.~Kovchegov,
  Phys.\ Rev.\ D {\bf 54}, 5463 (1996)
  doi:10.1103/PhysRevD.54.5463
  [hep-ph/9605446].


\bibitem{Lappi:2007ku} 
  T.~Lappi,
  Eur.\ Phys.\ J.\ C {\bf 55}, 285 (2008)
  doi:10.1140/epjc/s10052-008-0588-4
  [arXiv:0711.3039 [hep-ph]].


\bibitem{Song:2011hk} 
  H.~Song, S.~A.~Bass, U.~Heinz, T.~Hirano and C.~Shen,
  Phys.\ Rev.\ C {\bf 83}, 054910 (2011)
  Erratum: [Phys.\ Rev.\ C {\bf 86}, 059903 (2012)]
  doi:10.1103/PhysRevC.83.054910, 10.1103/PhysRevC.86.059903
  [arXiv:1101.4638 [nucl-th]].


\bibitem{Martinez:2012tu} 
  M.~Martinez, R.~Ryblewski and M.~Strickland,
  Phys.\ Rev.\ C {\bf 85}, 064913 (2012)
  doi:10.1103/PhysRevC.85.064913
  [arXiv:1204.1473 [nucl-th]].
  
\bibitem{Martinez:2010sd} 
  M.~Martinez and M.~Strickland,
  Nucl.\ Phys.\ A {\bf 856}, 68 (2011)
  doi:10.1016/j.nuclphysa.2011.02.003
  [arXiv:1011.3056 [nucl-th]].
  
\bibitem{Strickland:2016ezq} 
  M.~Strickland,
  EPJ Web Conf.\  {\bf 137}, 07026 (2017)
  doi:10.1051/epjconf/201713707026
  [arXiv:1611.05056 [nucl-th]].
  
\bibitem{Alqahtani:2017tnq} 
  M.~Alqahtani, M.~Nopoush, R.~Ryblewski and M.~Strickland,
  arXiv:1705.10191 [nucl-th].
  
\bibitem{Florkowski:2010cf} 
  W.~Florkowski and R.~Ryblewski,
  Phys.\ Rev.\ C {\bf 83}, 034907 (2011)
  doi:10.1103/PhysRevC.83.034907
  [arXiv:1007.0130 [nucl-th]].
  
\bibitem{Florkowski:2013lza} 
  W.~Florkowski, R.~Ryblewski and M.~Strickland,
  Nucl.\ Phys.\ A {\bf 916}, 249 (2013)
  doi:10.1016/j.nuclphysa.2013.08.004
  [arXiv:1304.0665 [nucl-th]].
  
\bibitem{Florkowski:2013lya} 
  W.~Florkowski, R.~Ryblewski and M.~Strickland,
  Phys.\ Rev.\ C {\bf 88}, 024903 (2013)
  doi:10.1103/PhysRevC.88.024903
  [arXiv:1305.7234 [nucl-th]].
  
\bibitem{Ryblewski:2012rr} 
  R.~Ryblewski and W.~Florkowski,
  Phys.\ Rev.\ C {\bf 85}, 064901 (2012)
  doi:10.1103/PhysRevC.85.064901
  [arXiv:1204.2624 [nucl-th]].
  
\end{thebibliography}
\end{document}